\documentclass[conference,compsoc]{IEEEtran}

\ifCLASSOPTIONcompsoc
  \usepackage[nocompress]{cite}
\else
  \usepackage{cite}
\fi

\usepackage{multirow}
\usepackage{setspace}
\usepackage[pdftex]{graphicx}
\graphicspath{{./img/}}
\usepackage{color,framed}
\usepackage{colortbl}
\usepackage{pifont}
\usepackage{xcolor}
\usepackage{mathrsfs}
\usepackage{tabularx}
\usepackage{amsfonts}
\usepackage{amsmath}
\usepackage{amsthm}
\usepackage{amssymb}
\usepackage{hyperref}
\usepackage{float}
\usepackage{url}
\usepackage{enumitem}
\usepackage{booktabs}
\usepackage{threeparttable}
\usepackage{bbding}
\usepackage{subfigure}
\usepackage{xspace}
\usepackage[linesnumbered,ruled,vlined,boxed]{algorithm2e}
\newcommand{\sysname}{\textsc{GCleaner}\xspace}
\definecolor{lightblue}{rgb}{0.75,0.9,0.9}

\begin{document}
%-------------------------------------------------------------------------------

%don't want date printed
\date{}

%\title{``No Matter What You Do!'': Mitigating Backdoor Attacks in \\Graph Neural Networks}

%\title{``No Matter What You Do'': Purifying Backdoored GNN Models \\ via Graph Unlearning}

\title{``\textit{No Matter What You Do}'': Purifying GNN Models via Backdoor Unlearning}

%\iffalse
\author{\IEEEauthorblockN{
Jiale Zhang\IEEEauthorrefmark{1},
Chengcheng Zhu\IEEEauthorrefmark{1},
Bosen Rao\IEEEauthorrefmark{1},
Hao Sui\IEEEauthorrefmark{2}, 
Xiaobing Sun\IEEEauthorrefmark{1},\\
Bing Chen\IEEEauthorrefmark{2},
Chunyi Zhou\IEEEauthorrefmark{3}, and
Shouling Ji\IEEEauthorrefmark{3}
}
\IEEEauthorblockA{\IEEEauthorrefmark{1}Yangzhou University}
\IEEEauthorblockA{\IEEEauthorrefmark{2}Nanjing University of Aeronautics and Astronautics}
\IEEEauthorblockA{\IEEEauthorrefmark{3}Zhejiang University}
}
%\fi

%for single author (just remove % characters)
% \author{
% {\rm Jiale Zhang}\\
% Yangzhou University
% \and
% {\rm Chengcheng Zhu }\\
% Yangzhou University
% \and
% {\rm Bosen Rao }\\
% Yangzhou University
% \and
% {\rm Hao Sui }\\
% Nanjing University of Aeronautics and Astronautics
% \and
% {\rm Xiaobing Sun }\\
% Yangzhou University
% \and
% {\rm Bing Chen }\\
% Nanjing University of Aeronautics and Astronautics
% \and
% {\rm Chunyi Zhou }\\
% Zhejiang University
% \and
% {\rm Shouling Ji }\\
% Zhejiang University
% % copy the following lines to add more authors
% % \and
% % {\rm Name}\\
% %Name Institution
% } % end author

\maketitle

%-------------------------------------------------------------------------------
\begin{abstract}
%-------------------------------------------------------------------------------
Graph Neural Networks (GNNs) fusion of local structure and node features through message passing enables powerful representation learning, driving their wide application across diverse fields. Recent studies have exposed that GNNs are vulnerable to several adversarial attacks, among which backdoor attack is one of the toughest. Similar to Deep Neural Networks (DNNs), backdoor attacks in GNNs lie in the fact that the attacker modifies a portion of graph data by embedding triggers and enforces the model to learn the trigger feature during the model training process. Despite the massive prior backdoor defense works on DNNs, defending against backdoor attacks in GNNs is largely unexplored, severely hindering the widespread application of GNNs in real-world tasks.

To bridge this gap, we present \sysname, the first backdoor mitigation method on GNNs. \sysname can mitigate the presence of the backdoor logic within backdoored GNNs by reversing the backdoor learning procedure, aiming to restore the model performance to a level similar to that is directly trained on the original clean dataset. To achieve this objective, we ask: \textit{How to recover universal and hard backdoor triggers in GNNs? How to unlearn the backdoor trigger feature while maintaining the model performance?} We conduct the graph trigger recovery via the explanation method to identify optimal trigger locations, facilitating the search of universal and hard backdoor triggers in the feature space of the backdoored model through maximal similarity. Subsequently, we introduce the backdoor unlearning mechanism, which combines knowledge distillation and gradient-based explainable knowledge for fine-grained backdoor erasure. Extensive experimental evaluations on four benchmark datasets demonstrate that \sysname can reduce the backdoor attack success rate to 10$\%$ with only $1\%$ of clean data, and has almost negligible degradation in model performance, which far outperforms the state-of-the-art (SOTA) defense methods. %Our code is available at \url{https://github.com/Graph-Axis/GCleaner}.
\end{abstract}

%-------------------------------------------------------------------------------
\section{Introduction}
%-------------------------------------------------------------------------------

Graph-structured data has become ubiquitous in various domains, including social networks \cite{s1}, mobile payment networks \cite{m1,m2}, and credit networks \cite{C1}. For instance, in social networks people are the nodes and their friendships constitute the edges. While in financial transactions, the nodes and edges could be people and their money transactions. Graphs have demonstrated a remarkable capacity to represent both objects and the diverse interactions between them. With the abstraction via graphs, many real-world problems that are related to networks and communities can be cast into a unified framework and solved by exploiting its underlying rich and deep mathematical theory as well as tremendously efficient computational techniques \cite{F1,F2}. Graph Neural Networks (GNNs) have recently emerged as highly successful models for processing graph-structured data \cite{GCN,GAT,GIN,GSA}. They learn effective graph representations by employing message passing strategies that recursively aggregate features from neighboring nodes. GNNs have surpassed traditional machine learning techniques and become the prevailing approach for various graph mining tasks.

\begin{figure}[h]
\centering
\includegraphics[width=1\linewidth]{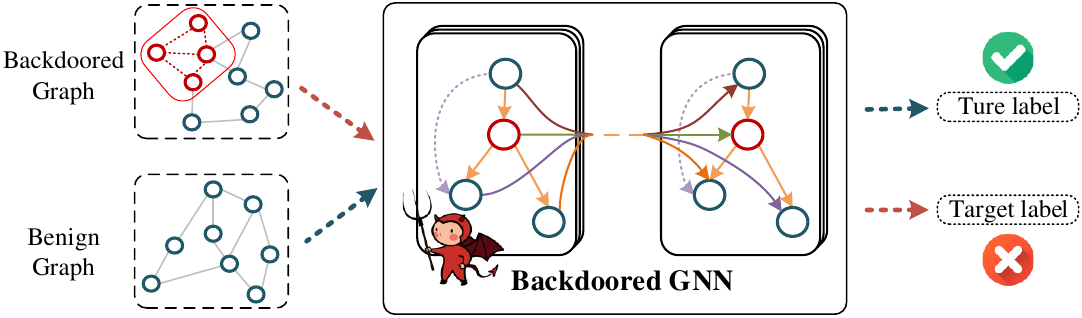}
\caption{Illustration of backdoor attack on GNNs.}
\label{F1}
\end{figure}

Despite the impressive performance of numerous GNNs in real-world tasks, concerns regarding their potential security issues have emerged. Studies have exposed the vulnerability of GNNs to adversarial attacks, which involve backdoor attacks \cite{NguyenThanh2023PoisoningGNNbased,Jiang2022CamouflagedPoisoning}, inference attacks \cite{Zhang2022InferenceAttacks,Wang2022GroupProperty,Conti2022LabelonlyMembership}, and reconstruction attacks \cite{Shen2022ModelStealing,Wu2022ModelExtraction}. Among them, backdoor attacks are emerging and becoming one of the most serious security threats to GNNs \cite{badsub,GTA,EXPBA,EXP2,motifBA,chen2022general,transferable,clean}. As shown in Figure \ref{F1}, the attacker seeks to insert a concealed backdoor into the targeted GNN model. The compromised GNN appears authentic when handling benign graphs. However, upon encountering an input marked with an attacker-defined "trigger", the victim GNN exhibits malicious behavior, such as misclassifying the manipulated input into a particular class. Existing research on graph backdoor attacks primarily focuses on the design of backdoor triggers, which can be roughly categorized into fixed triggers \cite{badsub,clean} and optimizable triggers \cite{GTA,transferable}. Furthermore, to enhance the effectiveness and stealthiness of backdoors, investigations have also been conducted on the injection locations of backdoor triggers \cite{EXPBA,EXP2,motifBA}.

Extensive prior work has been conducted on backdoor defense in computer vision (CV) and natural language processing (NLP) domains \cite{Gu2019BadNetsEvaluating,Li2021InvisibleBackdoor,Liu2019ABSScanning,Wang2019NeuralCleanse,Weber2023RABProvable,Kumari2023BayBFedBayesian,zhang2024flpurifier,Gong2023RedeemMyself,li2021anti}. Although some work proposed in these domains can be applied directly to the graph domain, their effectiveness is limited due to their disregard for the topological information inherent in graph data. %Unfortunately, the exploration of backdoor defense specifically tailored to GNNs is still in its early stages. 
To the best of our knowledge, only a few preliminary studies have focused on backdoor defense on graphs \cite{Detect,Detect1}. These approaches primarily concentrate on backdoor detection, utilizing activation statistics or model properties. They aim to identify potential backdoor examples within training or test data or to determine the presence of backdoors in models. However, in practical scenarios, there is an increasing reliance on pre-trained models or models trained by third parties, which introduces a significant risk, as model publishers can easily embed backdoors within these models \cite{contras, li2021anti}. In dealing with these pre-trained models, it is hard to detect backdoors at the original dataset level due to the inaccessibility of original datasets. Although some methods can determine whether a model contains backdoors, simply discarding a compromised model is not always a viable solution, as obtaining a pre-trained model often involves significant costs and efforts.  In summary, \textit{while detection plays a crucial role in identifying potential risks, it is insufficient as the backdoored model still requires purification}. This highlights the urgent need for a backdoor mitigation approach tailored specifically for graph data to alleviate the impact of backdoors in poisoned models.

\textbf{Our work.} In this work, we present \sysname, the first backdoor mitigation method for GNNs, which aims to reverse the backdoor learning procedure to achieve backdoor unlearning. Specifically, \sysname primarily consists of two key modules: trigger recovery and backdoor unlearning, where the former searches for potential triggers within the backdoored model, while the latter utilizes the recovered triggers to perform unlearning, effectively mitigating the backdoor. In the first phase, our observations revealed a consistent pattern: the model would predict the attacker-specified label whenever a trigger is present in any input. %That is, there is a specific trigger pattern that causes the backdoored model to output highly similar features. 
Utilizing this characteristic, we attempt to find a universal and hard trigger. Considering the node features and topological characteristics of graph data, we use explanation algorithm to obtain the node importance matrix of graphs, replacing insignificant nodes with potential trigger subgraphs, hoping that different graphs will output highly similar embeddings after embedding the trigger subgraph. Through this process, we continuously optimize the trigger subgraph to obtain a hard and universal trigger. In the backdoor unlearning stage, to achieve the erasure of backdoor logic while maintaining model performance, we implement unlearning through knowledge distillation. Specifically, we employ the fine-tuned backdoored model as the teacher model and the original backdoored model as the student model, with clean samples and trigger-embedded samples as inputs, it enforces that the backdoored model could forget the backdoor trigger features and clear the backdoor logic. To achieve more fine-grained unlearning, we further introduce a gradient-based graph explainable method to alleviate the damage of backdoor unlearning on normal information.

We evaluate our \sysname on four general datasets. Specifically, we transfer the backdoor mitigation method from the CV domain and implement the backdoor defense algorithm proposed in the existing work on graph backdoor attacks. Our experimental results show that with only $1\%$ of clean data, \sysname can reduce ASR to 10$\%$, with almost negligible degradation in model performance, far surpassing the baseline methods. Moreover, we also conduct numerous experiments to better understand each module.

In summary, our contributions lie in the following aspects:

\begin{itemize}
    \item To our best knowledge, this work presents the first study on the backdoor mitigation of GNNs. Correspondingly, we propose \sysname, a simple yet effective and fine-grained method for backdoor mitigation by reversing the attack process via backdoor unlearning.
    \item We conduct the graph trigger recovery via the explanation method to find universal and hard backdoor triggers. Subsequently, we introduce the backdoor unlearning mechanism, which combines knowledge distillation and gradient-based explainable knowledge for fine-grained backdoor erasure.
    \item We perform extensive experiments of \sysname on four general datasets: Bitcoin, Fingerprint, AIDS, and COLLAB. Evaluation results demonstrate that \sysname significantly surpasses the performance of existing backdoor defenses and also appears effective under various settings. 
\end{itemize}

\section{Background}
\subsection{Graph Neural Networks}
Given a graph $G = (V, E, X)$, where $V = \{v_1, v_2, \ldots, v_N\}$ signifies the set of nodes, $E$ represents the set of edges,  $X = \mathbb{R}^{N \times d}$ denotes the node attribute matrix, and $A \in \mathbb{R}^{N \times N}$ is the adjacency matrix. For any two nodes $v_i, v_j \in V$, if $(v_i, v_j) \in E$, it implies the presence of an edge between  $v_i$ with $v_j$ and $A_{ij} = 1$; otherwise, $A_{ij} = 0$. Here,  $N = |V|$ and $d$ denote the number of nodes and node feature dimensions.

The goal of graph learning is to acquire effective node representations by leveraging both structural information and node feature information through message propagation. One of the most prevalent models in graph learning is the Graph Convolutional Neural Network (GCN). Following the prior work \cite{GCN}, we define the graph convolutional layer to be 
\begin{equation}
	\begin{split}
     Z^{(l)}=\sigma(\tilde{D}^{-\frac{1}{2}}\tilde{A}\tilde{D}^{-\frac{1}{2}}Z^{(l-1)}W^{(l)}),
	\end{split}
	\label{EQ1}
\end{equation}
where $Z^l$ is the convolutional activations at the $l$-th layer, and initially, $Z^{0}=X$. $\tilde{A}=A+I$ is the adjacency matrix with added self-connections where $I\in\mathbb{R}^{n\times n}$ is the identity matrix and $\tilde{D}=D+I$ is the degree matrix of the graph. $\sigma(\cdot)$ is the element-wise nonlinear activation function such as $ReLU(\cdot)=\max(0,\cdot)$ and $W^{(l)}$ is the trainable parameter for the $l$-th GCN layers. Building upon the fundamental concept of GCN, numerous variants have been proposed to optimize and enhance the performance of GNNs from various perspectives. Most GNNs operate with a neighborhood aggregation strategy, which can be generalized by the following formula:
\begin{equation}
	\begin{split}
     Z^{(l)} = \text{\textbf{AGGERGATE}}(A,Z^{(l-1)}; W^{(l)}),
	\end{split}
	\label{EQ2}
\end{equation}
where $\text{\textbf{AGGERGATE}}$ function depends on the adjacency matrix $A$ and adopts different strategies in various GNNs.

The representation (i.e., embedding) of graphs, incorporating both the structure and node attributes, is learned by GNNs to facilitate diverse classification tasks.  Specifically, for Node Classification, given a graph where a subset of nodes is labeled as $V_{L}\subsetneq V$, with labels from $Y=\{y_{1},y_{2},...,y_{C}\}$, the node representations (normalized by a $\text{\textbf{SOFTMAX}}$ function) from the GNN, denoted as $f:V\rightarrow Y$, are employed to predict the labels of the remaining unlabeled nodes. In the graph classification tasks, a graph dataset $\mathcal{G}$ comprises $N$ graphs $\{(G_{1},y_{1}),(G_{2},y_{2}),...,(G_{N},y_{N})\}$, where $G_{i}$ is the $i$-th graph and $y_{i}$ is one of the C labels in the label space $Y=\{y_{1},y_{2},...,y_{C}\}$.The GNN model for graph classification trained by labeled graphs is regarded as $f:\mathcal{G}\rightarrow Y$. This model predicts the labels of graphs through graph-level embedding that is generated by pooling the node embeddings using the $\text{\textbf{READOUT}}$ function.

\subsection{Explainability Tools for GNNs}
Explanation techniques for deep models aim to delve into the underlying relationships that contribute to their predictions. Due to the remarkable ability to model graph-structured data, GNNs have gained increasing popularity, leading to the emergence of advanced GNN operations that enhance performance across diverse tasks. This growing trend has fueled extensive research on interpreting GNNs, yielding notable progress in this domain \cite{yuan2020xgnn,pope2019explainability,wang2020causal,luo2020parameterized,ying2019gnnexplainer}. In accordance with the types of explanations they provide, various techniques can be classified into two primary classes: instance-level \cite{pope2019explainability,wang2020causal,luo2020parameterized,ying2019gnnexplainer} and model-level \cite{yuan2020xgnn}. In this paper, we primarily focus on introducing instance-level methods to facilitate our task. Specifically, instance-level methods offer input-dependent explanations for each input graph, which explains the decision-making process of deep models by identifying crucial input for its prediction. Furthermore, based on how the importance scores are obtained, instance-level methods could be categorized into gradients/features-based methods \cite{pope2019explainability} and perturbation-based methods \cite{wang2020causal,luo2020parameterized,ying2019gnnexplainer,yuan2021explainability}.

\textbf{GNNExplainer} \cite{ying2019gnnexplainer} is a perturbation-based method that utilizes mask optimization technique to learn soft masks for edges and node features, enabling the interpretation of prediction outcomes. The process begins with the random initialization of soft masks, which are treated as trainable variables. These masks are then combined with the original graph through element-wise multiplications. Subsequently, the masks are optimized by maximizing the mutual information between the predictions of the original graph and the predictions obtained from the newly combined graph. GNNExplainer is a model-agnostic approach that can explain the predictions of any GNN model on diverse graph learning tasks without requiring modifications to the underlying GNN architecture or the need for re-training. Due to these advantages, the primary focus of this paper is on the GNNExplainer method. Formally, given a subgraph G, represented as Exp in GNNExplainer, it returns a mask indicating the influence of nodes on predictions within the input graph data:
\begin{equation}
	\begin{split}
     M = \textit{\text{Exp}}(G, y).
	\end{split}
	\label{EQ3}
\end{equation}
Here, higher values in the mask represent greater contributions of the corresponding nodes to the prediction outcomes, while lower values indicate smaller or insignificant contributions.

\textbf{Grad-CAM} (Gradient-weighted Class Activation Mapping) \cite{selvaraju2017grad} is a gradients/features-based method that originated in computer vision and has been applied to explain the attention of DNNs toward different regions within input images. Similarly, it can also be used to explain GNNs by employing gradients as weights to combine various feature maps and generating heatmaps to visualize these attention regions \cite{pope2019explainability}. Specifically, it calculates gradients of the target prediction with respect to the final node embeddings. Then, it averages these gradients to determine the weight for each feature map.

\section{Threat Model}

\textbf{Attacker’s goals.}
Backdoor attacks aim to inject malicious functions as hidden neural trojans into the target model so that when the pre-defined patterns, which are considered triggers, are present, the backdoor would be activated and mislead the model to output attacker-desired predictions. %Fundamentally, backdoor attacks establish a strong connection between triggers and the attacker's target labels via end-to-end training. Depending on the level of control the attacker exerts over the entire training process, these attacks can be categorized as black-box, gray-box, or white-box attacks. In black-box attacks, the attacker poisons the data, while in gray-box attacks, a small amount of data is utilized to compromise the model. White-box attacks involve complete control over the data and the entire training process. 
In \sysname, attackers generally aim for two main objectives: 
\\ \emph{\textbf{Effectiveness}}: Backdoored GNN models should output targeted class desired by attackers on trigger-embedded graphs.
\\\emph{\textbf{Stealthiness}}: Backdoor attack should not influence the normal accuracy on benign graphs, which ensures the stealthy.

Formally, suppose we have a clean GNN model $F$  and a backdoor GNN model $F_0$. The two objectives of the attack can be defined as:
\begin{equation}
	\begin{split}
\left.\left\{\begin{array}{l}F(G)=F_0(G)\\F(G^{\prime})=y_t\end{array}\right.\right.,
	\end{split}
	\label{EQ4}
\end{equation}
where $y_t$ represents the targeted attack class, $G$ represents a clean graph, and $G^{\prime}$ represents a graph with the trigger embedded in it. According to Eq. \ref{EQ4}, the first objective aims to ensure that the backdoored models exhibit similar behavior to benign models when operating on clean graphs. This objective makes the malicious models indistinguishable and difficult for defenders to detect, i.e., ensuring stealthiness. The second objective represents that the trojan model will predict the trigger-embedded graph as the targeted attack class, i.e., ensuring effectiveness.

\textbf{Attacker’s capability.}
Depending on different backgrounds and knowledge of the attack, we define attackers' capabilities into two cases: \ding{182}Attackers only poison a portion of the data and manipulate its labels. Subsequently, they release these poisoned data, hoping that users will incorporate them into their training data to obtain models, thus implanting a backdoor. \ding{183}Attackers can access a legitimate model and forge it into a backdoored model using an auxiliary dataset. %In \sysname, we assume attackers can fully control the training data and training process.%Regardless of the stage at which the attacker initiates the attack, the final goal of a backdoor attack is to obtain a model that contains a hidden backdoor.
In \sysname, regardless of the attacker's attack capability, their goal is to obtain a backdoored model. Thus, we assume attackers can fully control the training data and training process to manipulate a backdoored model from scratch.

\textbf{Defender’s capability and goal.}
Building upon previous studies \cite{Detect1,li2021neural,Wang2019NeuralCleanse,Gong2023RedeemMyself}, we constrain the defender's capabilities to enhance the fidelity of real-world scenario simulations: \ding{182}The defender remains oblivious to the identities of the tainted graphs or the attacker's target label. \ding{183} The defender can only access a limited portion of validation data or even public auxiliary data, and there are no specific requirements for the data distribution. %is granted to the defender. precluding possession of the entire training dataset. In an even more stringent scenario, the defender's resources are limited to a small subset sourced from publicly available auxiliary datasets. 
The goal of the defender is to decrease the success rate of backdoor attacks while maintaining the performance of the model on regular tasks. 

\begin{figure}[t]
\centering
\includegraphics[width=1\linewidth]{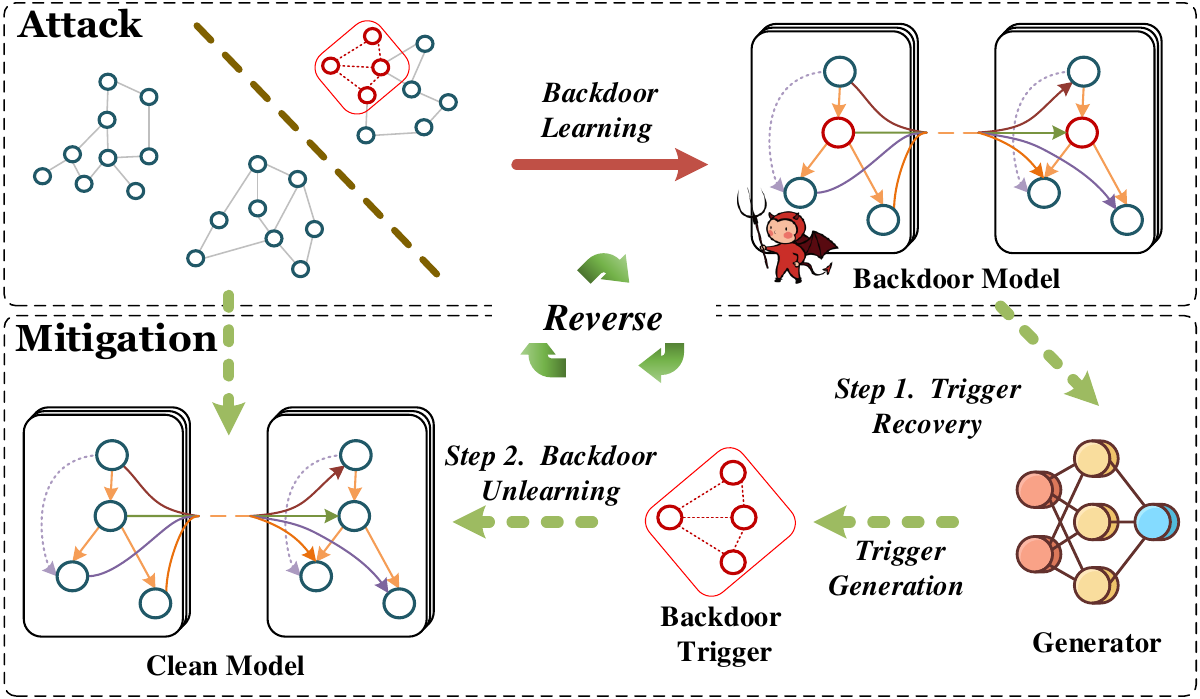}
%\vspace{-3mm}
\caption{Defense intuition of the proposed \sysname.}
\label{F2}
%\vspace{-3mm}
\end
{figure}

\begin{figure*}[h]
	\centering
	\includegraphics[width=0.95\linewidth]{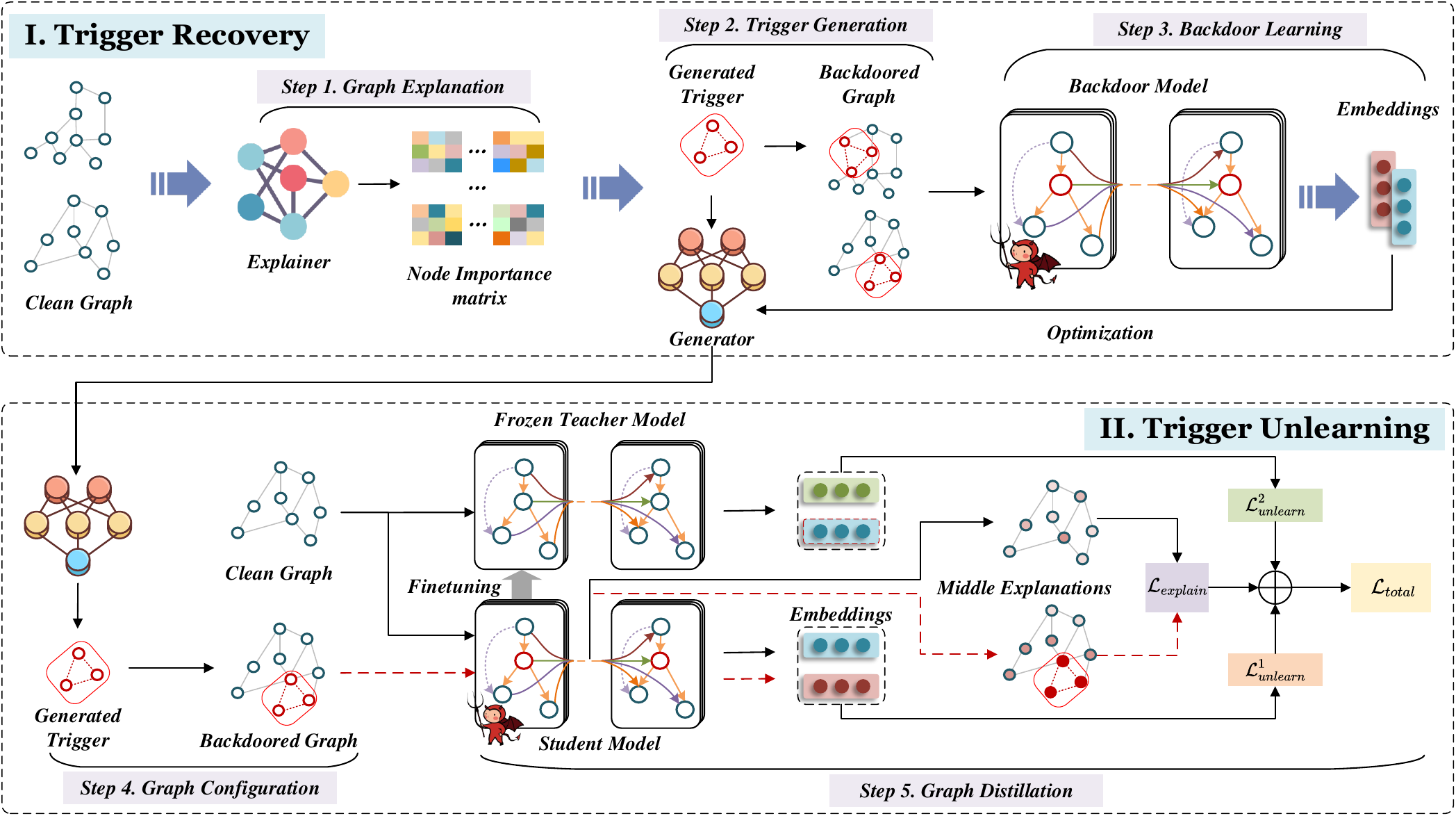}
% \vspace{-3mm}
	\caption{Framework of the proposed \sysname.} 
	\label{F3}
%	\vspace{-3mm}
\end{figure*}

\begin{figure}
	\centering
	\subfigure[Clean model.]{
		\begin{minipage}[t]{0.47\linewidth}
			\centering
			\includegraphics[width=1\linewidth]{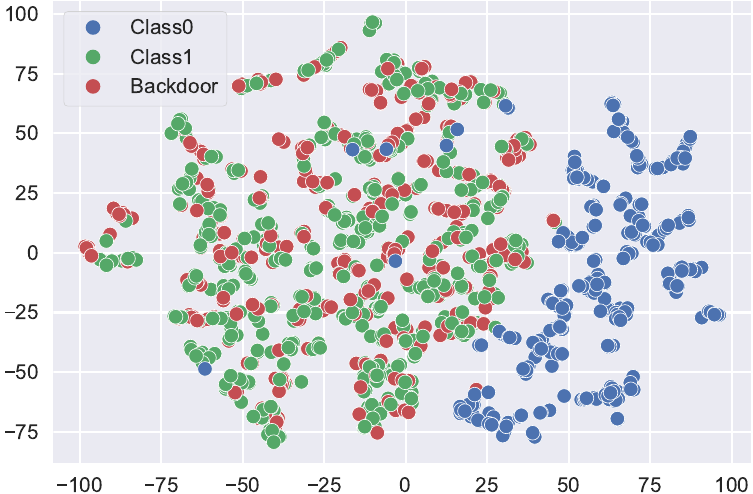}
			\label{F4a}
	\end{minipage}}
	\subfigure[Backdoored model.]{
		\begin{minipage}[t]{0.47\linewidth}
			\includegraphics[width=1\linewidth]{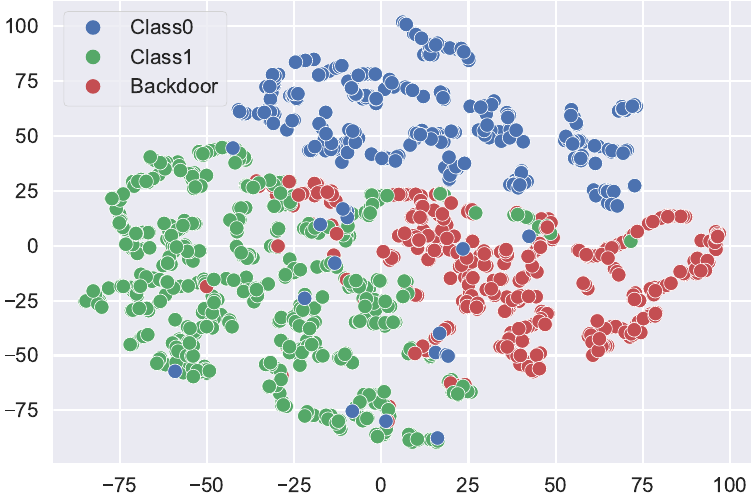}
			\label{F4b}
	\end{minipage}}%
% \vspace{-3mm}
	\caption{Embedding space distributions with trigger-embedding inputs.}
	\label{F4}
	%\vspace{-3mm}
\end{figure}

\section{Design of \sysname}
\label{Method}

\subsection{Defense Intuition and Challenges}
\label{Challenges}
As illustrated in Figure \ref{F2}, the core idea of a backdoor attack lies in the fact that the attacker modifies a portion of the data by embedding triggers and enforces the model to learn the trigger feature during the model training process. Consequently, the backdoored model classifies data embedded with triggers into the labels desired by the attacker. Our objective is to mitigate the presence of such backdoor logic within backdoored models, aiming to restore the performance of backdoored models to a level similar to that is directly trained on the original clean dataset.

Inspired by advancements in backdoor defense within the CV domain, a straightforward and intuitive approach involves \textit{reversing the process of backdoor learning}. Initially, the defender employs a reversal of the model prediction process, shifting from the conventional ``input$\rightarrow$output'' to ``output$\rightarrow$input'' sequence, where the input specifies the trigger patterns. In such ways, the polluted data will be recovered. Subsequently, the defender proceeds with a turning from the learning process to the unlearning process, which serves to eliminate the adverse impact of poisoned data on backdoored models. However, applying the aforementioned reversal process to defend against graph-based backdoors still poses the following challenges:

\textbf{\textit{Challenge 1: how to recover universal and hard backdoor triggers in GNNs?}}

Although backdoor trigger recovery has become a commonly used module in the backdoor defense process, trigger recovery tailored for GNNs has not yet been thoroughly explored. Compared with Euclidean structured data (such as images), graphs encompass both node features and topological features, rendering triggers for graph data more intricate and variable. Furthermore, the embedding location of triggers also significantly influences the feature representation of graph data. Therefore, there is a pressing need to devise a novel method for graph trigger recovery that can obtain universal and robust backdoor features.

\textbf{\textit{Challenge 2: how to unlearn the backdoor trigger feature while maintaining the model performance?}}

Since the problem of weak learners is easy to come out after unlearning the original data, especially under the condition of limited clean data, it is challenging to design a method that can efficiently forget malicious features and ensure the performance of the model. Furthermore, our objective is that the backdoored model selectively forgets the trigger subgraphs while retaining valuable information. Such fine-grained forgetting requirements further exacerbate the challenges.

\subsection{Overview}
To tackle the challenges highlighted in Section \ref{Challenges}, we have developed \sysname to address backdoor mitigation in GNNs. This section provides an overview of the comprehensive workflow of \sysname. An illustrative depiction of the system is presented in Figure \ref{F3}, emphasizing two primary modules: \ding{182}In the \emph{Trigger Recovery} module, \sysname employs explanation methods to identify optimal trigger locations, facilitating the search of universal and hard backdoor triggers in the feature space of the backdoored model through maximal similarity. \ding{183}In the \emph{Backdoor Unlearning} module, taking clean graphs and trigger-embedded graphs as input, \sysname employs the backdoored model as the student model, and the fine-tuned and frozen model as the teacher model. Through the teacher's guidance of the student's paradigm combined with intermediate explainable knowledge, \sysname achieves the backdoor unlearning while maintaining the model.

\subsection{Graph Trigger Recovery via Explanation Method}
Regardless of the targeted data type or model, the success of a backdoor attack is defined by the model's behavior: when the trigger appears on any data instance, the model consistently outputs the label specified by the attacker; conversely, for unpolluted inputs, the model operates normally. This phenomenon sheds light on the observation of the intrinsic mechanics of backdoor attacks.

\textbf{Observation 1:} \textit{Generally, attackers force the model to learn trigger features and establish a strong connection with the target label by embedding triggers in the sample and altering their labels to the ones desired by the attacker. Fundamentally, for clean and backdoored models, feature embeddings of the samples from the same class often gather together and form while the clusters formed from different classes have clear boundaries (the feature outputs vitalized by t-SNE are shown in Figure \ref{F4}). The trigger features will form a separate dense area and be close to the cluster of the target class.}

Our objective is to identify trigger features in the feature space, ensuring that samples with embedded trigger features can be mapped to the dense area. Naturally, we first draw inspiration from the trigger recovery method primarily applied in the visual domain, working by optimizing a trigger pattern, which can induce the targeted misclassification while having a small trigger size. We adapt this idea for graph trigger recovery. Firstly, we use the following equation to formalize the trigger injection:
\begin{equation}
\begin{split}
   \mathcal{F}(G,t,n)=G^{\prime},\\
    G^{\prime}=\text{Montage}(G,t),
\end{split}
\label{EQ5}
\end{equation}
where function $\mathcal{F}$ is employed to embed the trigger $t$ into an input graph $G$, resulting in a backdoored sample.  $G^{\prime}$ represents the trigger-embedding graph, and $n$ represents the number of nodes  in the trigger subgraph $t$. This embedding process is achieved through the use of the Montage function, which combines the trigger subgraph with the original graph to obtain the backdoored sample.

\begin{table}[]
 \caption{\small Similarity of inputs under different attacks.}
 \vspace{1mm}
  \centering
  \footnotesize
  \renewcommand\tabcolsep{6pt}
  \renewcommand\arraystretch{1}
\begin{tabular}{c|c|c}
\hline
                    & \textbf{Without (w/o) Trigger} & \textbf{With (w/) Trigger} \\ \hline
\textbf{Non-Attack} & 0.372                       & 0.415                      \\ \hline
\textbf{Sub-BA}     & 0.498                       & 0.874                      \\ \hline
\textbf{GTA}        & 0.471                       & 0.793                      \\ \hline
\textbf{Exp-BA}     & 0.425                       & 0.914                      \\ \hline
\textbf{Motif-BA}   & 0.417                       & 0.875                      \\ \hline
\end{tabular}
\label{T1}
%\vspace{-3mm}
\end{table}

In computer vision, the search for a minimal trigger pattern typically involves adding an optimizable trigger at arbitrary positions in clean samples. However, when dealing with graph data, certain distinctions arise: \ding{182}Graph data encompasses both node features and topological features, presenting a more complex structure than image data. \ding{183}The position of triggers within a graph significantly influences the feature representation. Despite these differences, both DNNs and GNNs share fundamental characteristics.

\textbf{Observation 2:} \textit{The model converges easily on backdoored data, indicating that it learns backdoored data much faster than learning with clean data. This rapid learning is attributed to the frequent appearance of triggers associated with a specific class throughout the training process. Consequently, the backdoored model takes a shortcut to learn general and hard trigger features. Any input embedded with such triggers will lead to the backdoored model producing highly similar feature embeddings (as shown in Table \ref{T1}).}

Building upon this observation, we attempt to achieve backdoor recovery by searching for a universal and hard backdoor feature in the feature space. In real-world scenarios, defenders are often unaware of the trigger size and embedding position set by attackers. However, leveraging the information propagation characteristics of GNN, regardless of the number of nodes contained in the trigger subgraph, after several rounds of message passing, they tend to aggregate onto a critical node, abstracting into a representative backdoor feature. It becomes inconsequential to determine the size of the trigger subgraph that needs to be recovered. Thus, we predefine triggers to contain $n$ nodes.

Furthermore, a backdoored model with a dense backdoor area requires only a slight perturbation to shift clean samples into this dense region. However, these minute perturbations are still distinct from trigger features. We aim to identify a representative trigger feature such that when it appears on any input, the model outputs similar embeddings. This implies that the backdoored model places greater emphasis on the backdoor feature than the original features of clean samples. To find such universal and hard trigger features, we employ an interpreter to yield the node importance matrix for clean samples. We then replace the least important nodes with a trigger subgraph. If the generated trigger nodes can cause the model prediction to deviate, that is, pay more attention to the backdoor nodes than the normal key nodes, it indicates the discovery of a universal and hard backdoor feature.

To reconstruct such trigger nodes, we train a trigger generator $f_{\theta_g}$ that takes node features as input. Specifically, we employ a Multi-Layer Perceptron (MLP) to simultaneously generate the node features and the structure of the trigger subgraph. The generation process is described as follows:
\begin{equation}
	\begin{split}
t = \left.\left\{\begin{array}{l} X_t=W_f\cdot\mathrm{MLP}(\mathbf{x}_i)
\\A_t=W_a\cdot\mathrm{MLP}(\mathbf{x}_i)\end{array}\right.\right.,
	\end{split}
	\label{EQ6}
\end{equation}
where $x_i$ represents the node features of $v_i$ in graph $G_i$, and $W_f$ and $W_a$ are learnable parameters for feature and structure generation, respectively. $X_t \in \mathbb{R}^{n \times d}$ denotes the synthetic features of the trigger nodes, where $n$ and $d$ represent the size of the generated trigger and the dimension of features. $A_t \in \mathbb{R}^{n \times n}$ represents the adjacency matrix of the generated trigger. Since real-world graphs are typically discrete, inspired by binary neural networks \cite{courbariaux2016binarized}, we binarize the continuous adjacency matrix $A_t$ during the forward computation, while the continuous values are used in backward propagation.

To optimize the generator $f_{\theta_g}$, we assume a clean dataset $\mathcal{G}_c$ and give one input $(G_{i},y_{i})$ from this dataset. We utilize an interpreter GNNExplainer to output its feature importance matrices $M_i = \textit{\text{Exp}}(G_{i}, y_{i})$, where the least important $n$ nodes' features are selected as input. Using generator $f_{\theta_g}$, we generate the trigger subgraph $t_i=(X_t^i,A_t^i)$ to replace the original nodes. Additionally, we employ cosine similarity to measure the similarity between two trigger-embedded samples. Formally,  given two inputs $G_{p}$ and $G_{q}$,  cosine similarity between their corresponding trigger-embedded samples can be expressed as follows:
\begin{equation}
	\begin{split}
\mathcal{L}_{p,q}(F,t,n)=-cos\Big(F\big(\mathcal{F}(G_p,t,n)\big),F\big(\mathcal{F}(G_q,t,n)\big)\Big).
	\end{split}
	\label{EQ7}
\end{equation}
To achieve high similarities among samples that approximate the search for the dense area in the embedding space of backdoored models,  samples a batch of inputs to stabilize the search process. The average of pair-wise similarity within a batch $B$ is computed as follows:
\begin{equation}
	\begin{split}
\min_{\theta_{g}}\mathcal{L}_{t}=\sum_{p\in B}\sum_{q\in B}\max(0,T-\mathcal{L}_{p,q}(F,t,n)),
	\end{split}
	\label{EQ8}
\end{equation}
 where $\mathcal{L}_{t}$ is used as the constraint during optimization, assuring that the samples stamped with the optimized trigger are in the dense area in the embedding space. $T$ is a threshold assuring the average similarity is high.
                      
\begin{table*}[t]
\centering
\caption{Dataset statistics.}
\vspace{1mm}
\small
\renewcommand\tabcolsep{8pt}
	\renewcommand\arraystretch{1}
\begin{tabular}{ccccccc}
 \hline \hline
\textbf{Datasets} & \textbf{\# Graphs} & \textbf{Avg.\# Nodes} & \textbf{Avg.\# Edges} & \textbf{\# Classes} & \textbf{\# Graphs[Class]}     & \textbf{\# Target Label} \\ \hline 
Bitcoin      & 1174              & 14.63                 & 14.18                 & 2                  & 845[0], 329[1] & 1                       \\
COLLAB            & 5000              & 73.49                & 2457.78              & 3                  & 2600[0],775[1],1625[2]       & 1                       \\ 
AIDS              & 2000              & 15.69                & 16.2                 & 2                  & 400[0],1600[1]               & 1                       \\
Fingerprint       & 1661              & 8.15                 & 6.81                 & 4                  & 538[0], 517[1],109[2],497[3] & 1                       \\
\hline \hline 
\end{tabular}
\label{T2}
\end{table*}
 
\subsection{Backdoor Unlearning via Graph Distillation}

Given the recovered trigger patterns, the next step of \sysname is to erase the trigger feature through machine unlearning. Considering Challenge 2, to efficiently forget the malicious feature utilizing limited clean data, we employ the idea of knowledge distillation to result in the unlearned model by selective knowledge transfer to the student model with a light teacher-student framework. 

Specifically, we first use clean data to fine-tune the backdoored model, and the fine-tuned backdoored model will be frozen and applied as the teacher model. The original backdoored model is used as a student model.  The unlearning objective is to remove the information about backdoor triggers while retaining the useful information in the student model, thereby purifying the backdoored model.  To achieve this goal, we sample clean data  $(G,y)$ from $\mathcal{G}_c$ and construct the poisoned data $G^{\prime}$  by adding the recovered trigger into $G$. Both $G$ and $G^{\prime}$ are taken as the input of the student model and their embeddings push each other to break the backdoor features. Meanwhile, the teacher model takes $G$ as input. Since the inputs to the fine-tuned teacher model do not contain trigger features, the teacher model will output clean soft targets. Such clean and useful information from the teacher model is passed on to the student which helps the student to forget the trigger feature. Furthermore, the clean output distribution of the student model should be closely aligned with the teacher model. This serves as mitigation of the adverse impact of excessive backdoor unlearning on the model performance on normal samples. Formally, the unlearning objective can be defined as follows: 
\begin{equation}
\begin{split}
\mathcal{L}_{unlearn} =\overbrace{\left\|F_{purif}^{logits}(G^{\prime})-F_{purif}^{logits}(G)\right\|_{2}}^{\mathcal{L}_{unlearn}^1}\\+\underbrace{\left\|F_{purif}^{logits}(G)-F_{frozen}^{logits}(G)\right\|_{2}}_{\mathcal{L}_{unlearn}^2}.
\end{split}
\label{EQ9}
\end{equation}
 Here, $F_{frozen}^{logits}(\cdot)$ and $F_{purif}^{logits}(\cdot)$ represent the logits output of the frozen fine-tuned backdoored model and backdoored model required to be purified, respectively. $\left\|\cdot\right\|_{2}$ denotes the $l_2$ norm, which is used to calculate the distance between two logits.

However, due to the characteristics of the aggregation of neighbor node features and label propagation of GCN, the hard feature of the trigger subgraph is easy to pass on to the clean nodes, making the feature output of the entire graph appear malicious. Hence, guiding the student model merely based on the logits or probability distribution is insufficient for more precise unlearning, especially when it comes to backdoor trigger features. Furthermore, such rough unlearning may lead to the loss of valuable information, despite the incorporation of constraint terms aimed at restoring useful knowledge. To achieve fine-grain backdoor unlearning, we adopt the extension of CNN explainability methods to GNNs to calculate the loss of the intermediate layer \cite{pope2019explainability}. Specifically, let the $k$-th graph convolutional feature map at layer $l$ be defined as:
\begin{equation}
\begin{split}
    F_k^l(X,A)=\sigma(VF^{(l-1)}(X,A)W_k^l),
\end{split}
\label{EQ10}
\end{equation}
where $W_k^l$ denotes the $k$-th column of matrix $W^l$, and $V = \tilde{D}^{-\frac12}\tilde{A}\tilde{D}^{-\frac12}$. In this notation, for node $n$, the $k$-th feature at the $l$-th layer is denoted by $F_{k,n}^l$. Then, the GAP feature after the final convolutional layer $L$, is calculated as
\begin{equation}
\begin{split}
    e_k=\frac{1}{N}\sum_{n=1}^NF_{k,n}^L(X,A),
\end{split}
\label{EQ11}
\end{equation}
and the class score is calculated as $\begin{aligned}y^c=\sum_kw_k^ce_k\end{aligned}$. Using this notation, Grad-CAM’s class specific weights for class $c$ at layer $l$ and for feature $k$ are calculated by
\begin{equation}
\begin{split}
    \alpha_k^{l,c}=\frac{1}{N}\sum_{n=1}^N\frac{\partial y^c}{\partial F_{k,n}^l},
\end{split}
\label{EQ12}
\end{equation}
and the heat map calculated from layer $l$ is
\begin{equation}
\begin{split}
    L_{heat}^c[l,n]=\text{ReLU}(\sum_k\alpha_k^{l,c}F_{k,n}^l(X,A)).
\end{split}
\label{EQ13}
\end{equation}

We compute the heat map after each layer of graph convolution and calculate the distance between that of the trigger-embedding input and clean input to achieve immediate knowledge transfer. The overall loss is a combination of the unlearning loss and the heat map loss:
\begin{equation}
\begin{split}
   &\mathcal{L}_{total}=\mathcal{L}_{unlearn} + \\ &\lambda\cdot\underbrace{\sum_{l=1}^K||L_{heat}^c(F_{purif}^{l}(G^{\prime}))-L_{heat}^c(F_{purif}^{l}(X))||_{2}}_{\mathcal{L}_{explain}},
\end{split}
\label{EQ14}
\end{equation}
where $\lambda$ is a weight coefficient to balance the two loss terms.

\begin{table*}[t]
\centering
\caption{Performance of \sysname compared with baseline methods on four datasets.}
\vspace{1mm}
\small
\renewcommand\tabcolsep{9pt}
	\renewcommand\arraystretch{1}
\begin{tabular}{cc|cccccccccc}
\toprule[0.8pt]
                                       &                                    & \multicolumn{2}{c}{\textbf{Before}}                                       & \multicolumn{2}{c}{\textbf{ABL}}                                          & \multicolumn{2}{c}{\textbf{Prune}}                                        & \multicolumn{2}{c}{\textbf{\begin{tabular}[c]{@{}c@{}}Randomized\\ -Smoothing\end{tabular}}} & \multicolumn{2}{c}{\textbf{\textit{\sysname}}}                                     \\ \cline{3-12} 
\multirow{-2}{*}{\textbf{Datasets}}    & \multirow{-2}{*}{\textbf{Attacks}} & {\color[HTML]{333333} \textbf{ASR}} & {\color[HTML]{333333} \textbf{ACC}} & {\color[HTML]{333333} \textbf{ASR}} & {\color[HTML]{333333} \textbf{ACC}} & {\color[HTML]{333333} \textbf{ASR}} & {\color[HTML]{333333} \textbf{ACC}} & {\color[HTML]{333333} \textbf{ASR}}           & {\color[HTML]{333333} \textbf{ACC}}          & {\color[HTML]{333333} \textbf{ASR}} & {\color[HTML]{333333} \textbf{ACC}} \\ \hline
                                       & \textbf{Sub-BA}                    & 79.03                               & \cellcolor{lightblue}68.48                               & 25.32                               & 64.85                               & 41.53                               & 62.27                               & 29.11                                         & 64.85                                        & \cellcolor{lightblue}0.00                                   & 67.88                               \\
                                       & \textbf{GTA}                       & 91.20                                & \cellcolor{lightblue}69.60                                & 28.86                               & 66.73                               & 35.63                               & 59.90                                & 35.16                                         & 67.51                                        & \cellcolor{lightblue}3.44                                & 68.07                               \\
                                       & \textbf{Exp-BA}                    & 84.67                               & \cellcolor{lightblue}69.17                               & 13.47                               & 64.58                               & 18.77                               & 61.31                               & 28.97                                         & 66.92                                        & \cellcolor{lightblue}0.38                                & 67.92                               \\
\multirow{-4}{*}{\textbf{Bitcoin}}     & \textbf{Motif-BA}                  & 86.08                               & \cellcolor{lightblue}68.25                               & 16.94                               & 61.99                               & 28.85                               & 58.59                               & 24.15                                         & 63.39                                        & \cellcolor{lightblue}0.71                                & 67.35                               \\ \hline
                                       & \textbf{Sub-BA}                    & 77.49                               & \cellcolor{lightblue}79.52                               & 21.42                               & 75.86                               & 42.60                                & 64.20                                & 32.02                                         & 67.71                                        & \cellcolor{lightblue}2.46                                & 79.14                               \\
                                       & \textbf{GTA}                       & 83.05                               & \cellcolor{lightblue}78.70                                & 32.70                                & 80.84                               & 35.19                               & 70.78                               & 45.37                                         & 76.90                                         & \cellcolor{lightblue}1.98                                & 77.10                                \\
                                       & \textbf{Exp-BA}                    & 79.42                               & \cellcolor{lightblue}78.92                               & 27.41                               & 71.57                               & 43.21                               & 68.43                               & 28.41                                         & 74.59                                        & \cellcolor{lightblue}3.25                                & 77.54                               \\
\multirow{-4}{*}{\textbf{COLLAB}}      & \textbf{Motif-BA}                  & 81.46                               & \cellcolor{lightblue}79.71                               & 24.64                               & 70.79                               & 22.97                               & 66.64                               & 39.48                                         & 71.55                                        & \cellcolor{lightblue}4.93                                & 77.84                               \\ \hline
                                       & \textbf{Sub-BA}                    & 95.84                               & \cellcolor{lightblue}97.14                               & 46.94                               & 95.36                               & 91.55                               & 87.25                               & 65.30                                          & 91.79                                        & \cellcolor{lightblue}6.64                                & 95.37                               \\
                                       & \textbf{GTA}                       & 99.02                               & \cellcolor{lightblue}97.33                               & 55.10                                & 95.00                                  & 97.18                               & 84.33                               & 79.59                                         & 87.86                                        & \cellcolor{lightblue}2.04                                & 94.64                               \\
                                       & \textbf{Exp-BA}                    & 96.76                               & 95.14                               & 33.89                               & 95.12                               & 62.24                               & 88.17                               & 45.51                                         & 92.36                                        & \cellcolor{lightblue}6.81                                & \cellcolor{lightblue}96.12                               \\
\multirow{-4}{*}{\textbf{AIDS}}        & \textbf{Motif-BA}                  & 95.91                               & \cellcolor{lightblue}97.85                               & 20.41                               & 96.43                               & 95.31                               & 83.25                               & 69.39                                         & 89.64                                        & \cellcolor{lightblue}4.08                                & 95.36                               \\ \hline
                                       & \textbf{Sub-BA}                    & 94.31                               & \cellcolor{lightblue}60.26                               & 12.60                                & 50.17                               & 35.01                               & 46.65                               & 80.32                                         & 50.00                                           & \cellcolor{lightblue}6.85                                & 59.55                               \\
                                       & \textbf{GTA}                       & 100.00                                 & \cellcolor{lightblue}61.54                               & 10.52                               & 54.74                               & 37.70                                & 49.58                               & 67.14                                         & 54.67                                        & \cellcolor{lightblue}6.88                                & 60.97                               \\
                                       & \textbf{Exp-BA}                    & 97.87                               & \cellcolor{lightblue}60.69                               & 18.13                               & 51.79                               & 29.79                               & 50.86                               & 59.71                                         & 56.61                                        & \cellcolor{lightblue}3.12                                & 58.65                               \\
\multirow{-4}{*}{\textbf{Fingerprint}} & \textbf{Motif-BA}                  & 98.97                               & \cellcolor{lightblue}61.11                               & 16.58                               & 58.54                               & 48.42                               & 53.17                               & 69.27                                         & 53.41                                        & \cellcolor{lightblue}5.15                                & 60.19                               \\ \bottomrule[0.8pt]
\end{tabular}
\label{T3}
\end{table*}

\section{Evaluation}\label{Evaluation}
In this section, we evaluate \sysname from different perspectives. First, we compare its performance with potential defense from other domains or existing arts. Then, we measure the effectiveness of \sysname under various backdoor settings and investigate the mechanism behind \sysname in depth. Finally, we do ablation studies to find out how the modules influence the performance of \sysname.
\subsection{Experimental Setup}
\noindent\textbf{Graph Datasets.}
We evaluate our approach on four real-world datasets (including Bitcoin\footnote{https://www.kaggle.com/datasets/jkraak/bitcoin-price-dataset}, COLLAB\footnote{https://paperswithcode.com/dataset/collab}, AIDS\footnote{https://paperswithcode.com/dataset/aids}, and Fingerprint\footnote{https://github.com/robertvazan/fingerprint-datasets}), where the basic statistics are shown in Table \ref{T2} and Appendix \ref{A.1}.%including the graph numbers in the dataset, the average number of nodes per graph, the average number of edges per graph, the number of classes, the number of graphs in each class, and the target class for attacks. %More detailed introduction and settings are in Appendix \ref{A.1}.

\noindent\textbf{Attack Methods.}
We investigate four SOTA backdoor attack methods (see in Appendix \ref{A.2}): Subgraph-based Bakdoor (Sub-BA) \cite{badsub}, GTA \cite{GAT}, Explainability-based Attack (Exp-BA) \cite{EXPBA}, and Motif-Backdoor (Motif-BA) \cite{motifBA}. %We provide a brief description of each method and settings in Appendix \ref{A.2}.

\noindent\textbf{Models.}
To evaluate the defense performance of \sysname, we choose four popular GNN models, i.e., Graph Convolutional Networks (GCN) \cite{GCN}, Graph Attention Networks (GAT) \cite{GAT}, Graph Isomorphism Network (GIN) \cite{GIN}, and Graph Sample and Aggregate (GraphSAGE) \cite{GSA} with publicly available implementations as the target models. Note that GIN is the default model unless otherwise mentioned.

\noindent\textbf{Baseline.}
To the best of our knowledge, \sysname is the first backdoor mitigation method against GNNs. We thus transfer the anti-backdoor learning (ABL) \cite{li2021anti}, one of the SOTA backdoor defense methods in CV domains, to graph fields. Further, we explore two possible defense methods from existing articles about graph backdoor attacks: pruning and randomized smoothing. Moreover, for existing backdoor detection methods, namely, ED \cite{Detect} and XGBD \cite{Detect1} since their main task is to identify malicious samples, we calculate the ASR of models trained by backdoor-filtered samples for a fairer comparison (see in Appendix \ref{A.3}).

\noindent\textbf{Evaluation Metrics.}
We evaluate the performance of defense mechanisms with two metrics: \ding{182}attack success rate (ASR), which is the ratio of trigger-embedded inputs misclassified by the backdoored model as the target labels specified by attackers:
\begin{equation}
\begin{split}
    \text{Attack Success Rate (ASR)}=\frac{\#\text{successful attacks}}{\#\text{total trials}},
\end{split}
\label{EQ15}
\end{equation}
and \ding{183}the accuracy of the main classification task on normal samples (ACC). An effective defense method means significantly reducing the ASR while maintaining a high ACC.

\noindent\textbf{Implementation Details.}
We implement \sysname in Python using the PyTorch framework. Our experimental environment consists of 13th Gen Intel(R) Core(TM) i7-13700KF, NVIDIA GeForce RTX 4070 Ti, 32GiB memory, and Ubuntu 20.04 (OS). We split the data into training data and test data in a ratio of 8:2, and \sysname is assumed to be able to access 3$\%$ of the clean data randomly selected from the testing set. We run the trigger recovery for 20 epochs. For the unlearning process, we use the loss term $\lambda = 0.5$, batch size $B = 64$, SGD as optimizer with learning rate $\eta = 0.001$, and run for $E = 30$ epochs. Notably, in most cases, it takes only around 10 epochs to reduce the ASR to a very low level, indicating that although our method includes both recovery and unlearning steps, the computational overhead remains acceptable. For all the baseline attacks and defenses, we adopt the default hyperparameters recommended by the corresponding papers. Specifically, attacks have the common parameters: trigger size $t$ and injection ratio $\varphi$. Given a dataset $D_{train}$ with an average node number $N_{avg}$, the number of nodes in the subgraph trigger is equal to $N_{avg} \times t$. Unless otherwise mentioned, we adopt the backdoor injection ratio to $\varphi = 5\%$ and the trigger size $t$ to $20\%$. We test the performance of \sysname as well as other baselines five times and report the mean and standard deviation results to eliminate the effects of randomness.

\subsection{Experimental Results}\label{Effectiveness}
% To evaluate the effectiveness of \sysname, we repeated each experiment 10 times and reported their average statistical performance \cite{DBLP:conf/ccs/KleesRCW018}. We empirically set the threshold for each gadget chain to 15 gadgets. For each statically identified gadget chain, we limit the fuzzing campaign of \sysname to 120 seconds. We performed additional sensitivity analysis on these two hyperparameters for evaluation in \refappendix{Sensitivity}.

\subsubsection{Comparison}

\indent Table \ref{T3} illustrates the performance comparison of \sysname with three baseline methods across four datasets. In the ``Before'' column, the results of attacks without any defense mechanisms are displayed. The best results are highlighted in \colorbox{lightblue}{Blue} color. Overall, \sysname significantly outperforms the baseline methods across all datasets against four graph backdoor attacks, maintaining the ASR below 7$\%$ while experiencing only a slight decrease in ACC accuracy.

Specifically, the performance of ABL lags far behind its effectiveness in backdoor defense within the CV scenario. Our analysis attributes this primarily to its reliance on an intuitive detection of backdoor samples, where the decrease in loss for backdoor samples during training occurs at a faster rate than for normal samples. %In image classification tasks, pixel-level feature representations are commonly used, treating each pixel as a feature. Consequently, backdoor attackers can manipulate pixel values in images to initiate backdoor attacks, making these samples easier for the model to learn, leading to a faster loss decrease. In contrast, feature representations in graph tasks typically revolve around node and edge attributes, requiring attackers to modify attributes of nodes or edges. 
In image classification, where pixel-level features are used, attackers can easily alter pixel values to execute backdoor attacks, accelerating the model's learning of these samples. Conversely, in graph tasks, where feature representations focus on node and edge attributes, attackers must adjust these attributes, making backdoor attacks less straightforward. Due to the unique information propagation mechanisms in graphs, such triggers are less likely to be learned fast by the model during training. Additionally, ABL results in a significant decrease in backdoor ACC, indicating interdependencies among features in graphs, where forgetting trigger features solely through gradient ascent leads to the loss of normal features, thus impacting the model's regular performance.

\begin{figure}
	\centering
	\subfigure[Clean model.]{
		\begin{minipage}[t]{0.31\linewidth}
			\centering
			\includegraphics[width=1\linewidth]{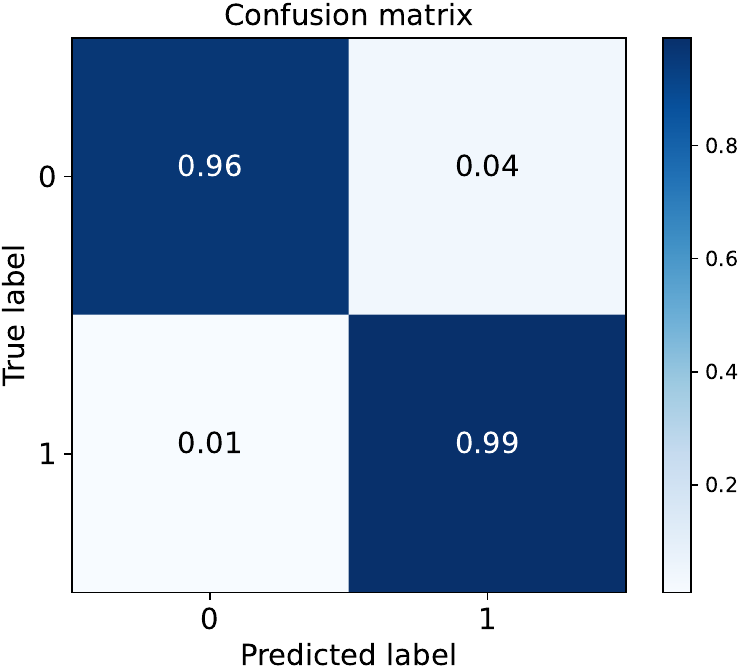}
			\label{F5a}
	\end{minipage}}
    \subfigure[backdoored model.]{
	\begin{minipage}[t]{0.31\linewidth}
			\includegraphics[width=1\linewidth]{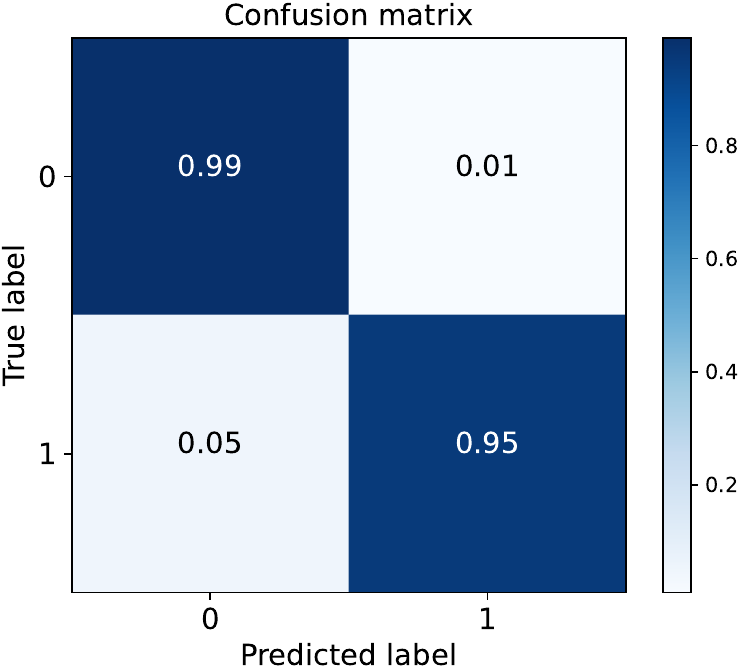}
			\label{F5b}
	\end{minipage}}%
	\subfigure[Unlearning model.]{
		\begin{minipage}[t]{0.31\linewidth}
			\includegraphics[width=1\linewidth]{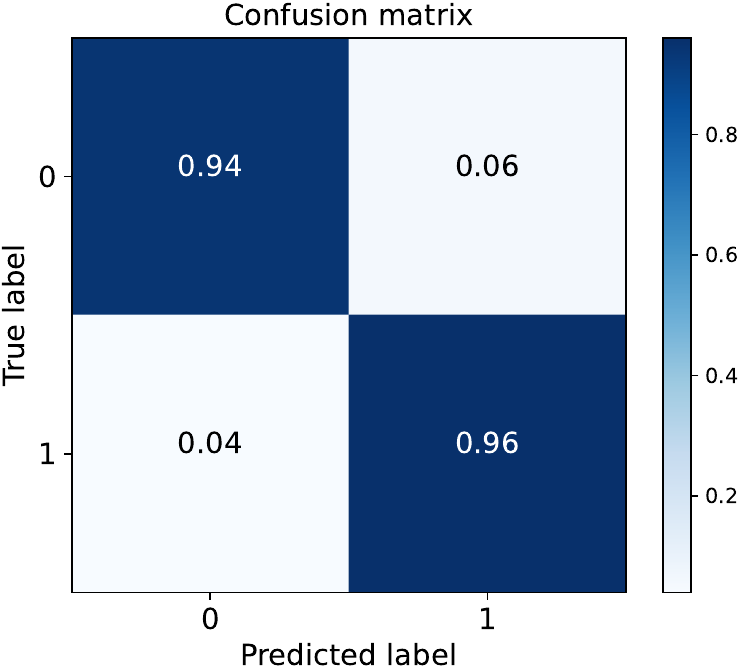}
			\label{F5c}
	\end{minipage}}%
	\caption{Classification accuracy for each class of different models.}
	\label{F5}
	%\vspace{-3mm}
\end{figure}

Prune identifies nodes with significantly low cosine similarity to the features of their connected nodes for pruning. However, this method is effective only in datasets that align with this feature. Moreover, attackers can easily circumvent this approach by strategically selecting trigger injection points or optimizing trigger features. While Randomized-Smoothing demonstrates certain effectiveness on select datasets, it sacrifices ACC to achieve better defense effectiveness. Balancing these aspects poses a challenging problem. In comparison, our method, through the recovery of universal and robust trigger features combined with fine-grained unlearning, is better suited for graph scenarios. Additionally, we visualize feature embeddings of both backdoored models and unlearning models for the feature level analysis (see in Appendix \ref{A.6}). 

Furthermore, we investigated the impact of unlearning on the classification accuracy of the target class. Since both the backdoor injecting and removal inevitably lead to the damage of normal features for these classes, these operations tend to decrease the accuracy of the attacker's target class. We present the results using confusion matrices as shown in Figure \ref{F5}. The experimental results indicate that although there is some decline in the classification accuracy of the target class, we significantly mitigate this issue by introducing graph heat maps. By focusing the unlearning process on backdoor features, we achieve fine-grained erasure of these features, greatly alleviating the problem.

%\vspace{-3mm}
\subsubsection{Effective of Trigger Recovery}

For our method, the effectiveness of backdoor recovery is foundational to the success of unlearning. However, in real-world scenarios, we do not have access to any information about the trigger, such as its injection method, location, or size. Therefore, we aim to identify a general trigger that can approximate the effect of the original one. This process is akin to searching for a perturbation that can cause the backdoored model to misclassify to the target class. %We introduce perturbations at the least important nodes to optimize a more representative, i.e., harder perturbation.
In this part, we validate the effectiveness of such a strategy. We compare the recovery effect of randomly adding triggers during the recovery process with that of using different explanation methods (GNNExplainer \cite{ying2019gnnexplainer}, PGExplainer \cite{luo2020parameterized} and SubgraphX \cite{yuan2021explainability}) to identify the least important locations for trigger injection. Note that we randomly add triggers to nodes outside of these critical subgraphs for explanation methods where the output identifies key subgraphs. Specifically, we measure the cosine similarity between the features of the backdoored model when embedding the original trigger and the recovered trigger samples to assess the trigger recovery effect. Table \ref{T4} presents the experimental results, where Recovery-ASR represents the backdoor success rate of the samples embedded with the recovered trigger, and Unlearn-ASR indicates the backdoor success rate after unlearning using the recovered trigger.

\begin{table}[]
 \caption{Trigger recovery effectiveness with different explanation methods.}
 \vspace{1mm}
  \centering
  \small
  \renewcommand\tabcolsep{2.5pt}
  \renewcommand\arraystretch{1}
\begin{tabular}{ccccc}
\hline
Method                        & Attack   & Similarity & \begin{tabular}[c]{@{}c@{}}ASR\\ (Recovery)\end{tabular} & \begin{tabular}[c]{@{}c@{}}ASR\\ (Unlearn)\end{tabular} \\ \hline
\multirow{4}{*}{Random}       & Sub-BA   & 0.685      & 95.76                                                    & 15.25                                                   \\
                              & GTA      & 0.814      & 100.00                                                   & 17.54                                                   \\
                              & Exp-BA   & 0.761      & 88.63                                                    & 19.14                                                   \\
                              & Motif-BA & 0.564      & 82.20                                                    & 30.78                                                   \\ \hline
\multirow{4}{*}{GNNExplainer} & Sub-BA   & 0.625      & 98.35                                                    & 6.64                                                    \\
                              & GTA      & 0.624      & 100.00                                                   & 2.04                                                    \\
                              & Exp-BA   & 0.687      & 97.73                                                    & 6.81                                                    \\
                              & Motif-BA & 0.636      & 79.66                                                    & 4.08                                                    \\ \hline
\multirow{4}{*}{PGExplainer}  & Sub-BA   & 0.691      & 100.00                                                   & 5.93                                                    \\
                              & GTA      & 0.557      & 100.00                                                   & 7.02                                                    \\
                              & Exp-BA   & 0.728      & 97.74                                                    & 4.39                                                    \\
                              & Motif-BA & 0.603      & 93.22                                                    & 5.13                                                    \\ \hline
\multirow{4}{*}{SubgraphX}    & Sub-BA   & 0.634      & 87.42                                                    & 7.32                                                    \\
                              & GTA      & 0.659      & 98.17                                                    & 2.18                                                    \\
                              & Exp-BA   & 0.713      & 91.35                                                    & 4.50                                                    \\
                              & Motif-BA & 0.602      & 94.08                                                    & 3.18                                                    \\ \hline
\end{tabular}
\label{T4}
\end{table}

\begin{figure*}[h]
	\centering
	\subfigure[ASR with Sub-BA method.]{
		\begin{minipage}[t]{0.235\linewidth}
			\centering
			\includegraphics[width=1\linewidth]{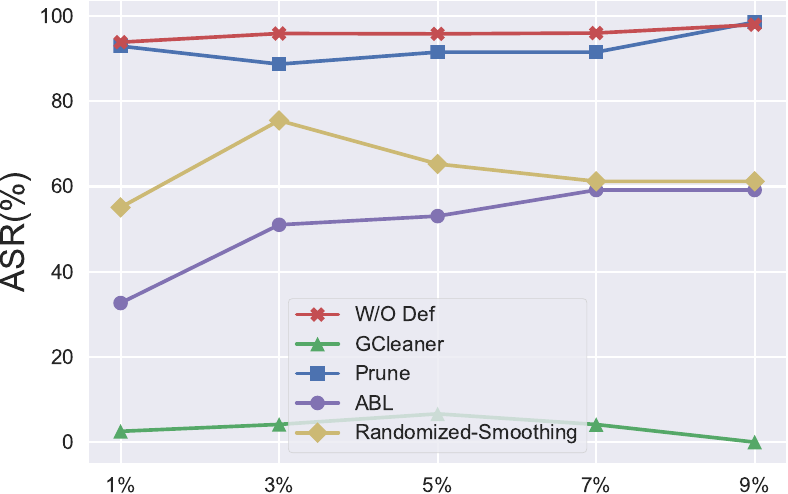}
			\label{F6a}
	\end{minipage}}
	\subfigure[ASR with GTA method.]{
		\begin{minipage}[t]{0.235\linewidth}
			\includegraphics[width=1\linewidth]{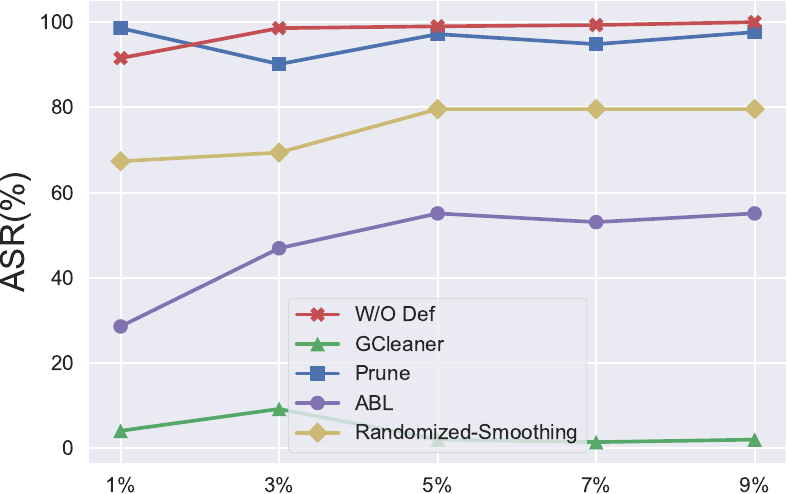}
			\label{F6b}
	\end{minipage}}
	\subfigure[ASR with Exp-BA method.]{
		\begin{minipage}[t]{0.235\linewidth}
			\centering
			\includegraphics[width=1\linewidth]{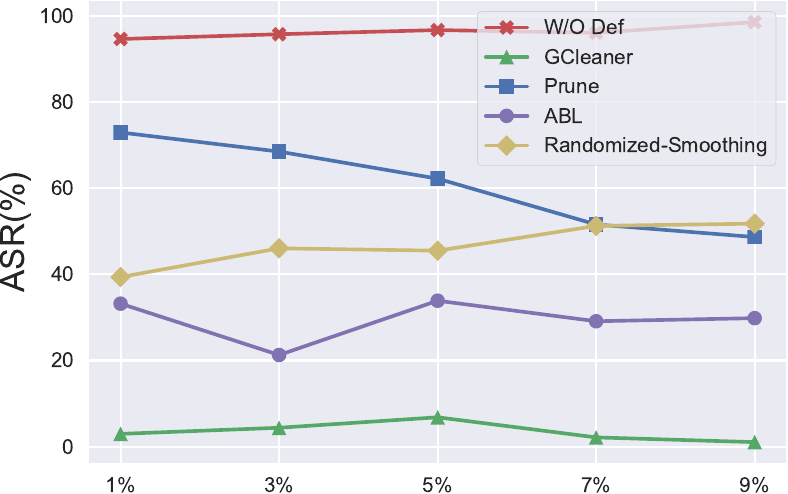}
			\label{F6c}
	\end{minipage}}
	\subfigure[ASR with Motif-BA method.]{
		\begin{minipage}[t]{0.235\linewidth}
			\includegraphics[width=1\linewidth]{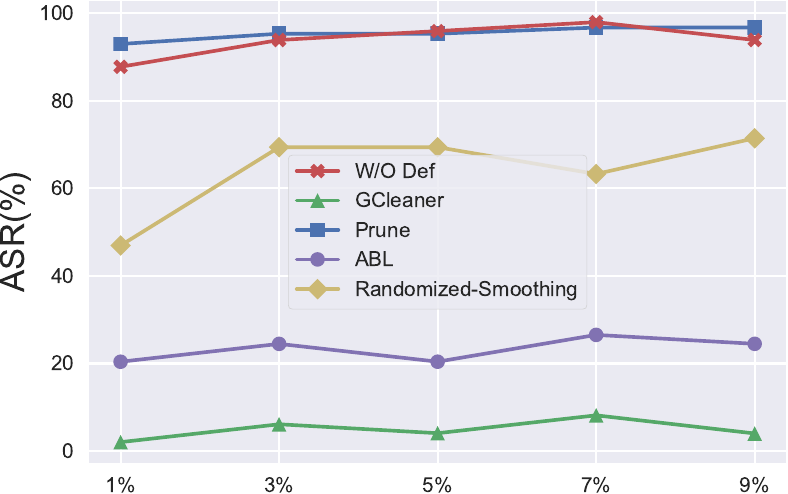}
			\label{F6d}
	\end{minipage}}%
 \\
 \subfigure[ACC with Sub-BA method.]{
		\begin{minipage}[t]{0.235\linewidth}
			\centering
			\includegraphics[width=1\linewidth]{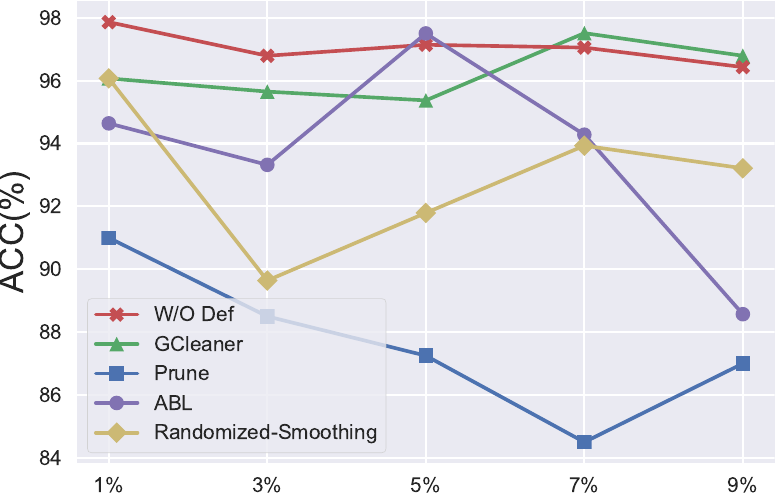}
			\label{F6e}
	\end{minipage}}
	\subfigure[ACC with GTA method.]{
		\begin{minipage}[t]{0.235\linewidth}
			\includegraphics[width=1\linewidth]{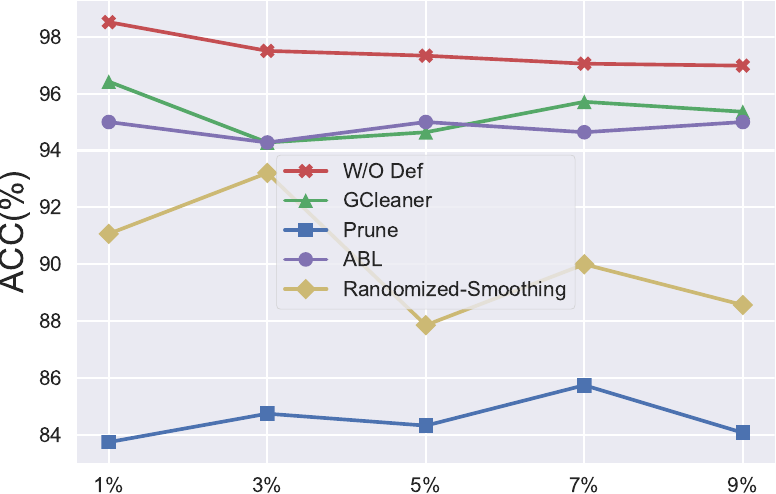}
			\label{F6f}
	\end{minipage}}
	\subfigure[ACC with Exp-BA method.]{
		\begin{minipage}[t]{0.235\linewidth}
			\centering
			\includegraphics[width=1\linewidth]{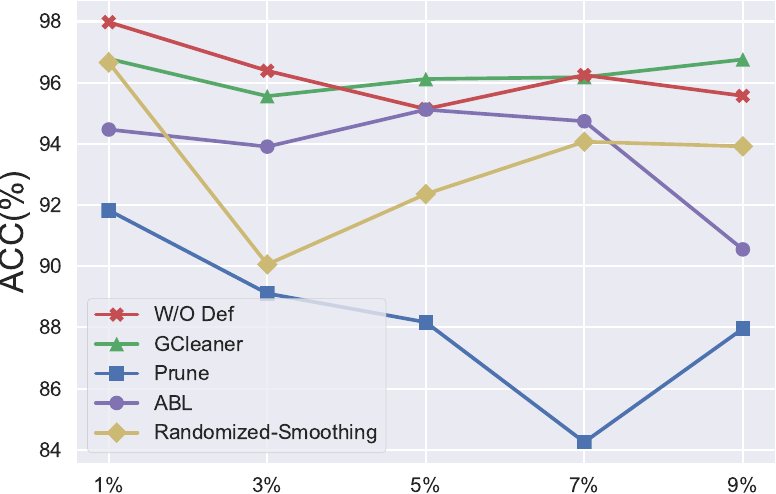}
			\label{F6g}
	\end{minipage}}
	\subfigure[ACC with Motif-BA method.]{
		\begin{minipage}[t]{0.235\linewidth}
			\includegraphics[width=1\linewidth]{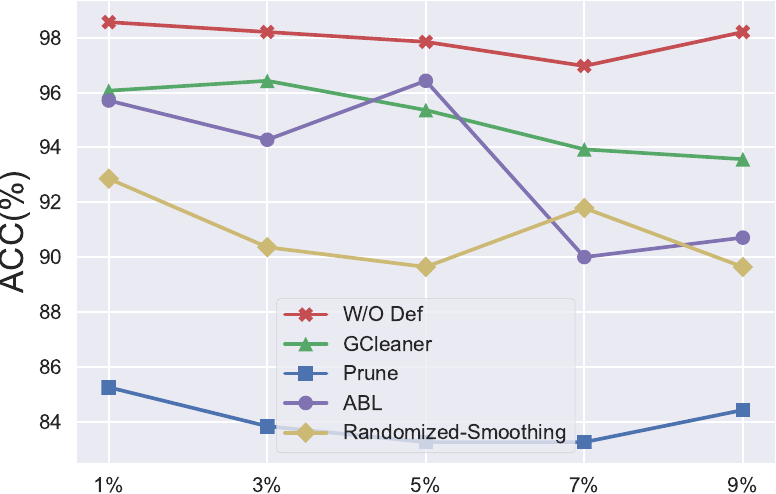}
			\label{F6h}
	\end{minipage}}%
	\caption{Impact of injection ratio with four SOTA methods.}
	\label{F6}
%	\vspace{-2mm}
\end{figure*}

\begin{figure*}
	\centering
	\subfigure[ASR on Bitcoin dataset.]{
		\begin{minipage}[t]{0.235\linewidth}
			\centering
			\includegraphics[width=1\linewidth]{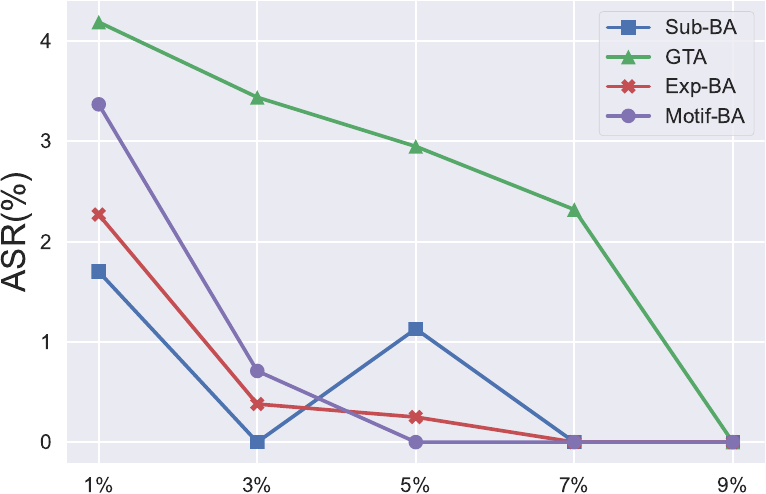}
			\label{F7a}
	\end{minipage}}
	\subfigure[ASR on Fingerprint dataset.]{
		\begin{minipage}[t]{0.235\linewidth}
			\includegraphics[width=1\linewidth]{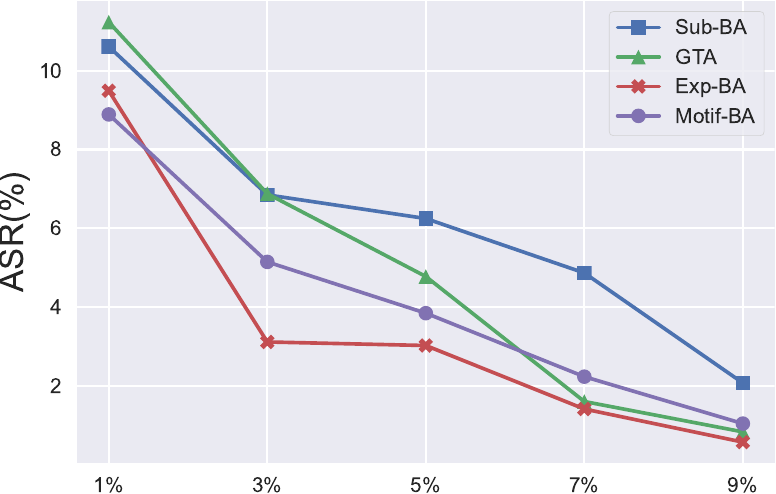}
			\label{F7b}
	\end{minipage}}
	\subfigure[ASR on AIDS dataset.]{
		\begin{minipage}[t]{0.235\linewidth}
			\centering
			\includegraphics[width=1\linewidth]{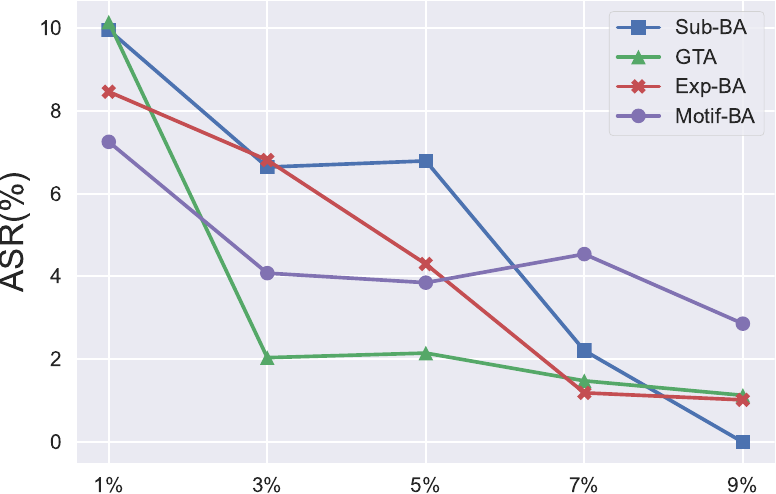}
			\label{F7c}
	\end{minipage}}
	\subfigure[ASR on COLLAB dataset.]{
		\begin{minipage}[t]{0.235\linewidth}
			\includegraphics[width=1\linewidth]{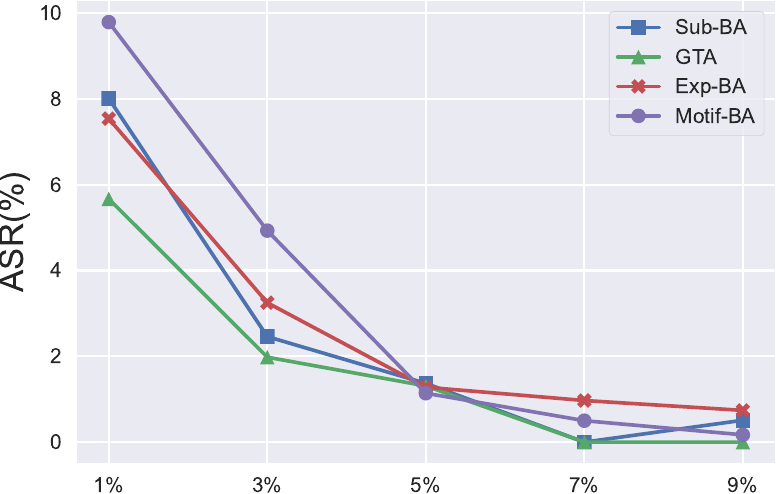}
			\label{F7d}
	\end{minipage}}%
 \\
 \subfigure[ACC on Bitcoin dataset.]{
		\begin{minipage}[t]{0.235\linewidth}
			\centering
			\includegraphics[width=1\linewidth]{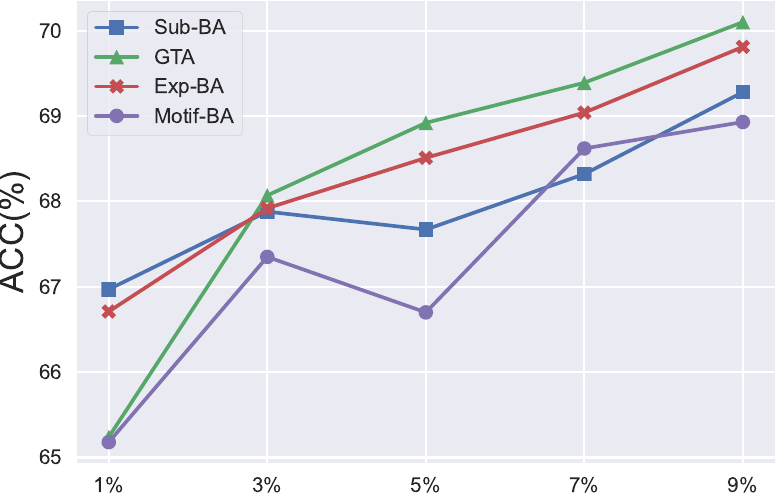}
			\label{F7e}
	\end{minipage}}
	\subfigure[ACC on Fingerprint dataset.]{
		\begin{minipage}[t]{0.235\linewidth}
			\includegraphics[width=1\linewidth]{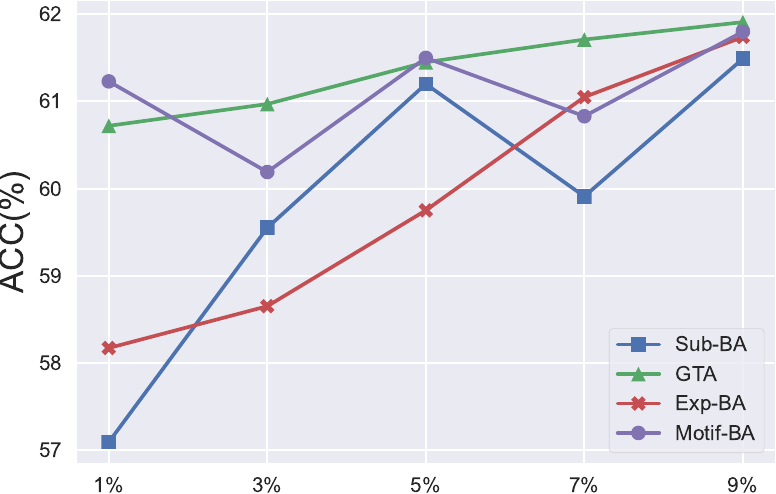}
			\label{F7f}
	\end{minipage}}
	\subfigure[ACC on AIDS dataset.]{
		\begin{minipage}[t]{0.235\linewidth}
			\centering
			\includegraphics[width=1\linewidth]{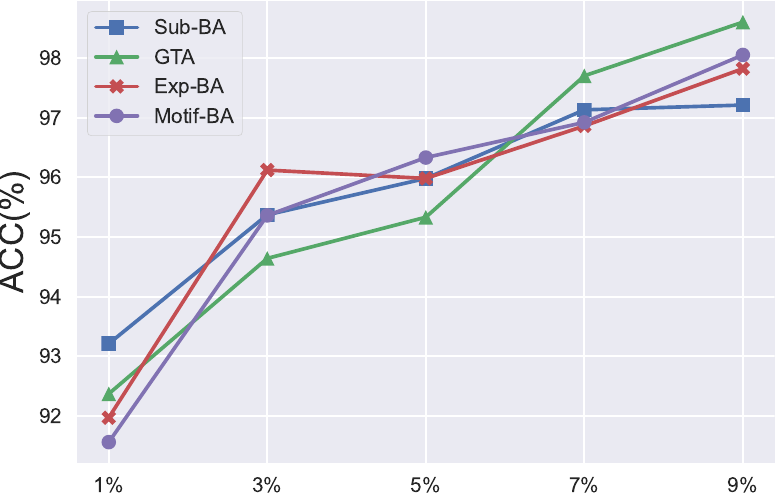}
			\label{F7g}
	\end{minipage}}
	\subfigure[ACC on COLLAB dataset.]{
		\begin{minipage}[t]{0.235\linewidth}
			\includegraphics[width=1\linewidth]{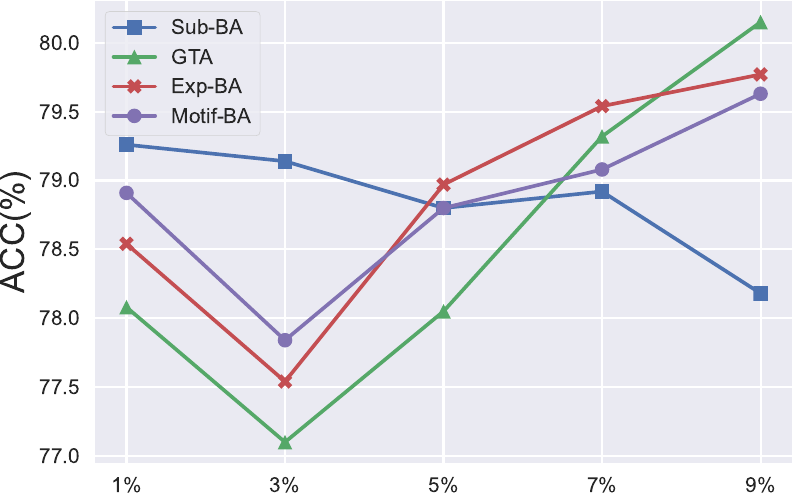}
			\label{F7h}
	\end{minipage}}%
	\caption{Impact of holding rate on four datasets.}
	\label{F7}
	%\vspace{-2mm}
\end{figure*}

The experimental results show that the backdoor triggers recovered at different injection locations can achieve a relatively high backdoor success rate. However, during the unlearning process, the random approach is unsatisfactory. This aligns with our expectations, as the random method may optimize a naturally occurring perturbation that causes the model’s prediction to deviate but is not representative. Additionally, we found that different explanation methods achieved similar results, which may be due to the relative similarity in the identification of unimportant nodes by different explanations. Furthermore, we investigate the impact of the predefined number of recovered triggers $N$ in Appendix \ref{A.7}.

\begin{figure*}[h]
	\centering
	\subfigure[Results on Bitcoin dataset.]{
		\begin{minipage}[t]{0.23\linewidth}
			\centering
			\includegraphics[width=1\linewidth]{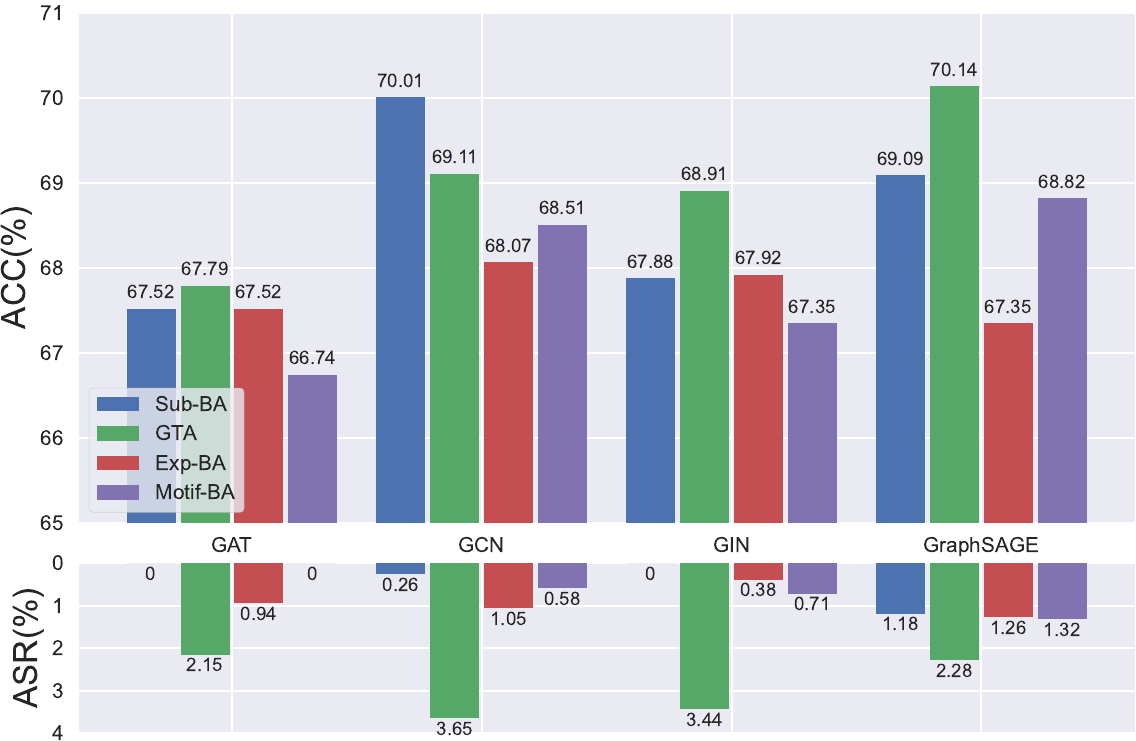}
			\label{F8a}
	\end{minipage}}
	\subfigure[Results on Fingerprint dataset.]{
		\begin{minipage}[t]{0.23\linewidth}
			\includegraphics[width=1\linewidth]{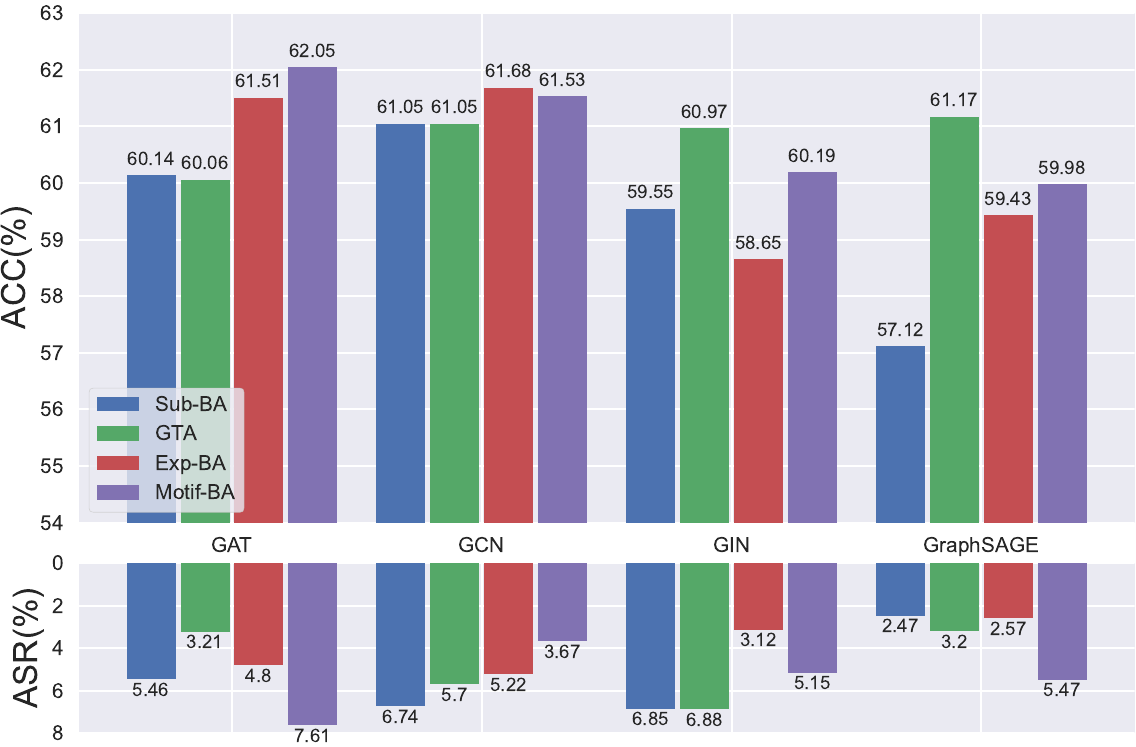}
			\label{F8b}
	\end{minipage}}
	\subfigure[Results on AIDS dataset.]{
		\begin{minipage}[t]{0.23\linewidth}
			\centering
			\includegraphics[width=1\linewidth]{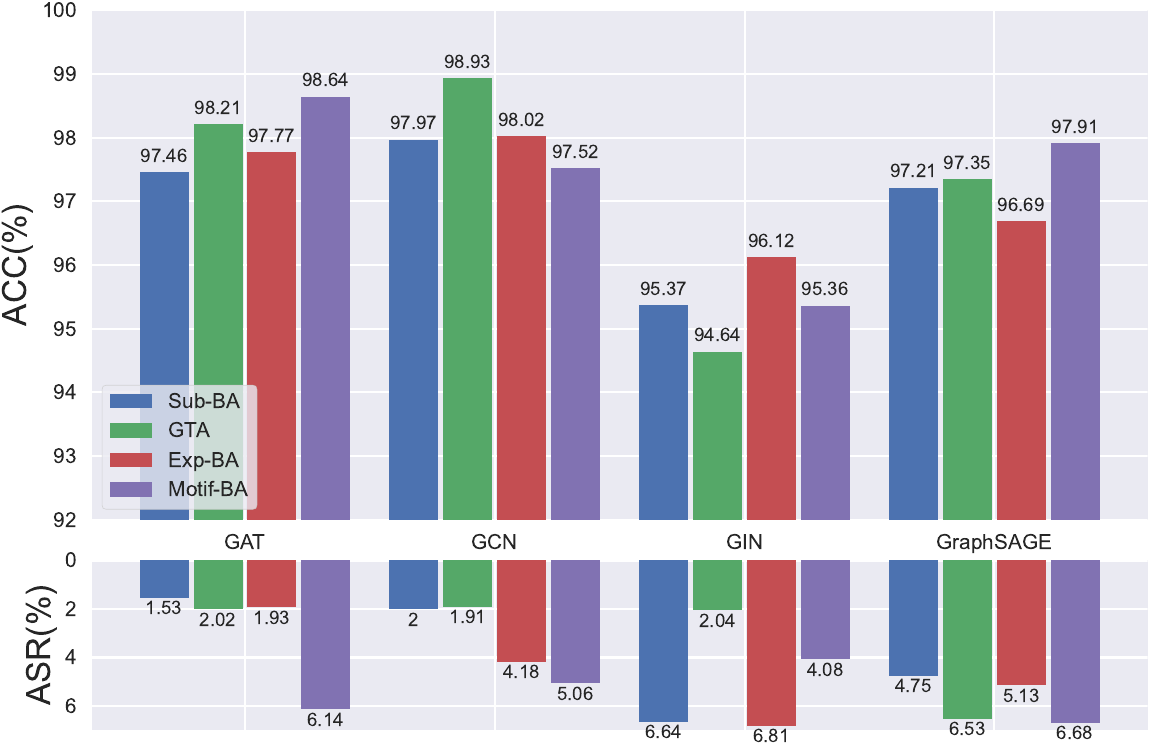}
			\label{F8c}
	\end{minipage}}
	\subfigure[Results on COLLAB dataset.]{
		\begin{minipage}[t]{0.23\linewidth}
			\includegraphics[width=1\linewidth]{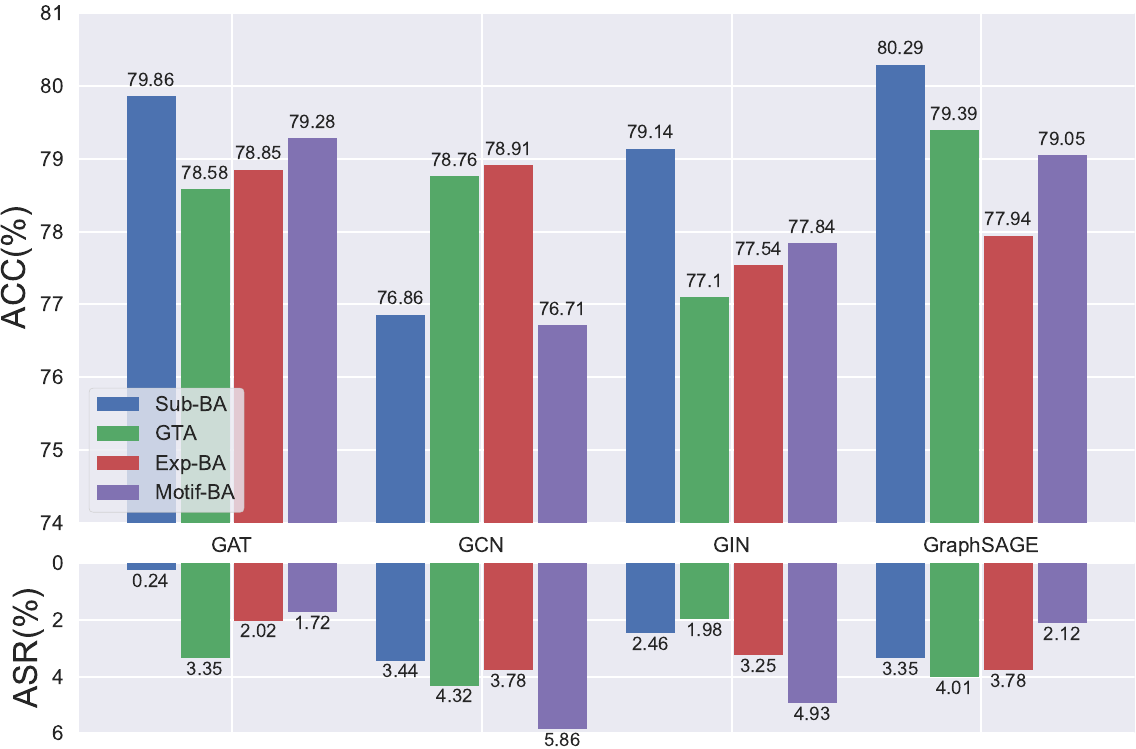}
			\label{F8d}
	\end{minipage}}%
	\caption{Impact of model types on four datasets.}
	\label{F8}
	%\vspace{-2mm}
\end{figure*}

%\vspace{-3mm}
\subsubsection{Impact of Backdoor Attacks Settings}

\
\newline\indent\textbf{Impact of Injection Ratio.} 
Due to the defender's limited exposure to only the backdoored model, the proportion of backdoor injections in the training dataset remains unknown. To verify the robustness of our method against different proportions of backdoor attacks, we compared the resilience of various defense methods against four attacks with varying backdoor injection rates on the AIDS dataset. The results are illustrated in Figure \ref{F6}, where the backdoor injection rates range from 1$\%$ to 9$\%$ in 2$\%$ increments.

Intuitively, without any defense mechanisms in place, the ASR increases with higher levels of backdoor injections, leading to a trade-off between injecting more poisoned data and a decline in the accuracy of the primary task. Overall, a higher proportion of backdoor injections escalates the difficulty of backdoor defense. Notably, our method demonstrates superior defense performance when the backdoor injection rate surpasses 7$\%$. Our analysis attributes this to the heightened success rate of backdoor attacks at higher injection rates, thereby facilitating easier and more accurate recovery of a superior backdoor trigger through \sysname. %This, in turn, enables deeper unlearning of backdoor features, resulting in enhanced backdoor defense efficacy.
That is, our method exhibits excellent performance across various backdoor injection scenarios, showcasing its resilience and effectiveness in mitigating backdoor attacks. Regarding the impact of trigger size $t$ on \sysname, please refer to the Appendix \ref{A.4}.

%\vspace{-3mm}
\subsubsection{Further Understanding of \sysname}
\
\newline \indent\textbf{Impact of Holding Rate.} As mentioned in Section \ref{Method}, a clean dataset is essential for the process of trigger subgraph recovery and backdoor feature unlearning. In the experimental setting, defenders extract a certain proportion of clean data from the test set, representing the clean data holding ratio, which is the ratio of the number of clean data samples to the total data size. Considering the challenge of acquiring clean datasets in most real-world applications, we maintain the clean data ratio below 10$\%$ and evaluate the robustness of our method within the range of 1$\%$ to 10$\%$, as shown in Figure \ref{F7}.

As expected, \sysname demonstrates better defense performance as more clean data samples are available. This is attributed to its performance being nearly equivalent to retraining a stronger model in the presence of abundant clean data points. While this is practically unfeasible, limited data can also help in maintaining the ASR of backdoor attacks below 10$\%$ across four datasets with only 1$\%$ of clean data.

\begin{table}[t]
 \caption{Performance of \sysname on alternative datasets.}
 \vspace{1mm}
  \centering
  \small
  \renewcommand\tabcolsep{7pt}
  \renewcommand\arraystretch{1}
\begin{tabular}{c|c|cc|cc}
\hline
\multirow{2}{*}{\textbf{Setting}}                                                              & \multirow{2}{*}{\textbf{Attack}} & \multicolumn{2}{c|}{\textbf{Origin}}                                                                                                                                        & \multicolumn{2}{c}{\textbf{Auxiliary}} \\ \cline{3-6} 
                                                                                               &                                  & \textbf{ASR}                                                                       & \textbf{ACC}                                                                           & \textbf{ASR}       & \textbf{ACC}      \\ \hline
\multirow{4}{*}{\textbf{\begin{tabular}[c]{@{}c@{}}NCI1\\ -\textgreater{}AIDS\end{tabular}}}   & \textbf{Sub-BA}                  & \multirow{8}{*}{\begin{tabular}[c]{@{}c@{}}6.64\\ 2.04\\ 6.81\\ 4.08\end{tabular}} & \multirow{8}{*}{\begin{tabular}[c]{@{}c@{}}95.37\\ 94.64\\ 96.12\\ 95.36\end{tabular}} & 11.18              & 91.56             \\
                                                                                               & \textbf{GTA}                     &                                                                                    &                                                                                        & 12.28              & 90.36             \\
                                                                                               & \textbf{Exp-BA}                  &                                                                                    &                                                                                        & 9.26               & 88.33             \\
                                                                                               & \textbf{Motif-BA}                &                                                                                    &                                                                                        & 9.32               & 92.92             \\ \cline{1-2} \cline{5-6} 
\multirow{4}{*}{\textbf{\begin{tabular}[c]{@{}c@{}}COLLAB\\ -\textgreater{}AIDS\end{tabular}}} & \textbf{Sub-BA}                  &                                                                                    &                                                                                        & 13.35              & 88.05             \\
                                                                                               & \textbf{GTA}                     &                                                                                    &                                                                                        & 11.01              & 89.39             \\
                                                                                               & \textbf{Exp-BA}                  &                                                                                    &                                                                                        & 10.78              & 91.94             \\
                                                                                               & \textbf{Motif-BA}                &                                                                                    &                                                                                        & 9.12               & 89.05             \\ \hline
\end{tabular}
  \label{T5}
  % \vspace{-3mm}
\end{table}

\begin{table}[h]
 \caption{\small Performance of \sysname with explanation methods.}
 \vspace{1mm}
  \centering
  \small
  \renewcommand\tabcolsep{6.5pt}
  \renewcommand\arraystretch{1.1}
\begin{tabular}{c|ccc}
\hline
\textbf{Attack}                     & \textbf{}             & \textbf{ASR}                                 & \textbf{ACC}            \\ \hline
                                    & \textbf{Non}          & 13.01                                        & 92.41                   \\
                                    & \textbf{GNNExplainer} & {16.95($\uparrow$\textcolor{red}{3.94})} & 93.51($\uparrow$\textcolor{green}{1.10})   \\
                                    & \textbf{IGradients}           & 6.78($\downarrow$\textcolor{green}{6.23})                       & 93.41($\uparrow$\textcolor{green}{1.00})   \\
                                    & \textbf{Guided BP}          & 9.32($\downarrow$\textcolor{green}{3.69})                       & 92.58($\uparrow$\textcolor{green}{0.17})   \\
\multirow{-5}{*}{\textbf{Sub-BA}}   & \textbf{Ours}         &\cellcolor{lightblue} 6.64($\downarrow$\textcolor{green}{6.37})                       & \cellcolor{lightblue}95.37($\uparrow$\textcolor{green}{2.96})   \\ \hline
                                    & \textbf{Non}          & 5.26                                         & 91.00                   \\
                                    & \textbf{GNNExplainer} & 8.77($\uparrow$\textcolor{red}{3.51})                                  & 92.57($\uparrow$\textcolor{green}{1.57})   \\
                                    & \textbf{IGradients}           & 3.62($\downarrow$\textcolor{green}{1.64})                       & 92.75($\uparrow$\textcolor{green}{1.75})   \\
                                    & \textbf{Guided BP}          & \cellcolor{lightblue}1.75($\downarrow$\textcolor{green}{3.51})                       & 87.57($\downarrow$\textcolor{red}{3.43}) \\
\multirow{-5}{*}{\textbf{GTA}}      & \textbf{Ours}         & 2.04($\downarrow$\textcolor{green}{3.22})                       & \cellcolor{lightblue}94.64($\uparrow$\textcolor{green}{3.64})   \\ \hline
                                    & \textbf{Non}          & 17.64                                        & 88.71                   \\
                                    & \textbf{GNNExplainer} & 9.47($\downarrow$\textcolor{green}{8.17})                       & 91.73($\uparrow$\textcolor{green}{3.02})   \\
                                    & \textbf{IGradients}           & 9.42($\downarrow$\textcolor{green}{8.22})                       & 93.67($\uparrow$\textcolor{green}{4.96})   \\
                                    & \textbf{Guided BP}          & 8.59($\downarrow$\textcolor{green}{9.05})                       & 90.92($\uparrow$\textcolor{green}{2.21})   \\
\multirow{-5}{*}{\textbf{Exp-BA}}   & \textbf{Ours}         & \cellcolor{lightblue}6.81($\downarrow$\textcolor{green}{10.83})                      & \cellcolor{lightblue}96.12($\uparrow$\textcolor{green}{7.41})   \\ \hline
                                    & \textbf{Non}          & 11.97                                        & 90.50                   \\
                                    & \textbf{GNNExplainer} & 8.55($\downarrow$\textcolor{green}{3.42})                       & 91.01($\uparrow$\textcolor{green}{0.51})   \\
                                    & \textbf{IGradients}           & \cellcolor{lightblue}2.13($\downarrow$\textcolor{green}{9.84})                       & 91.51($\uparrow$\textcolor{green}{1.01})   \\
                                    & \textbf{Guided BP}          & 5.98($\downarrow$\textcolor{green}{5.99})                       & 92.37($\uparrow$\textcolor{green}{1.87})   \\
\multirow{-5}{*}{\textbf{Motif-BA}} & \textbf{Ours}         & 4.08($\downarrow$\textcolor{green}{7.89})                       & \cellcolor{lightblue}95.36($\uparrow$\textcolor{green}{4.86})   \\ \hline
\end{tabular}
  \label{T6}
%  \vspace{-3mm}
\end{table}

In more extreme scenarios where no data from the original dataset is accessible, we resort to using alternative datasets, such as publicly available datasets, to aid in backdoor mitigation. Specifically, we select two additional datasets, NCI1 and COLLAB, with matching feature dimensions, to mitigate the backdoors introduced by four attack methods on the AIDS dataset. Experimental results in Table \ref{T5} demonstrate that by leveraging other auxiliary datasets, our method can also reduce the ASR to around 10$\%$. This indicates that the attack's effectiveness heavily relies on establishing a strong correlation between the injected trigger and the target embedding. Consequently, the process of trigger recovery essentially involves discovering a universal ``perturbation'' that biases the model, and we achieve backdoor defense through unlearning, making the model robust to such perturbations.

\textbf{Impact of Model Types.} %In order to ascertain the generalizability of our approach, 
We conducted a validation of our method across four widely recognized models: GAT, GCN, GIN, and GraphSAGE. As depicted in Figure \ref{F8}, the experimental outcomes reveal nuanced performance variations when exposed to diverse attacks across these distinct models. Despite encountering various forms of attacks and model intricacies, our approach consistently demonstrates effectiveness within acceptable margins. Notably, when subjected to diverse attack scenarios, the ASR for all four models remains consistently minimal, falling below 8$\%$. This finding underscores the model-agnostic nature of our approach.%, highlighting its robust versatility across varying model architectures.

\begin{figure}
\centering
\includegraphics[width=0.95\linewidth]{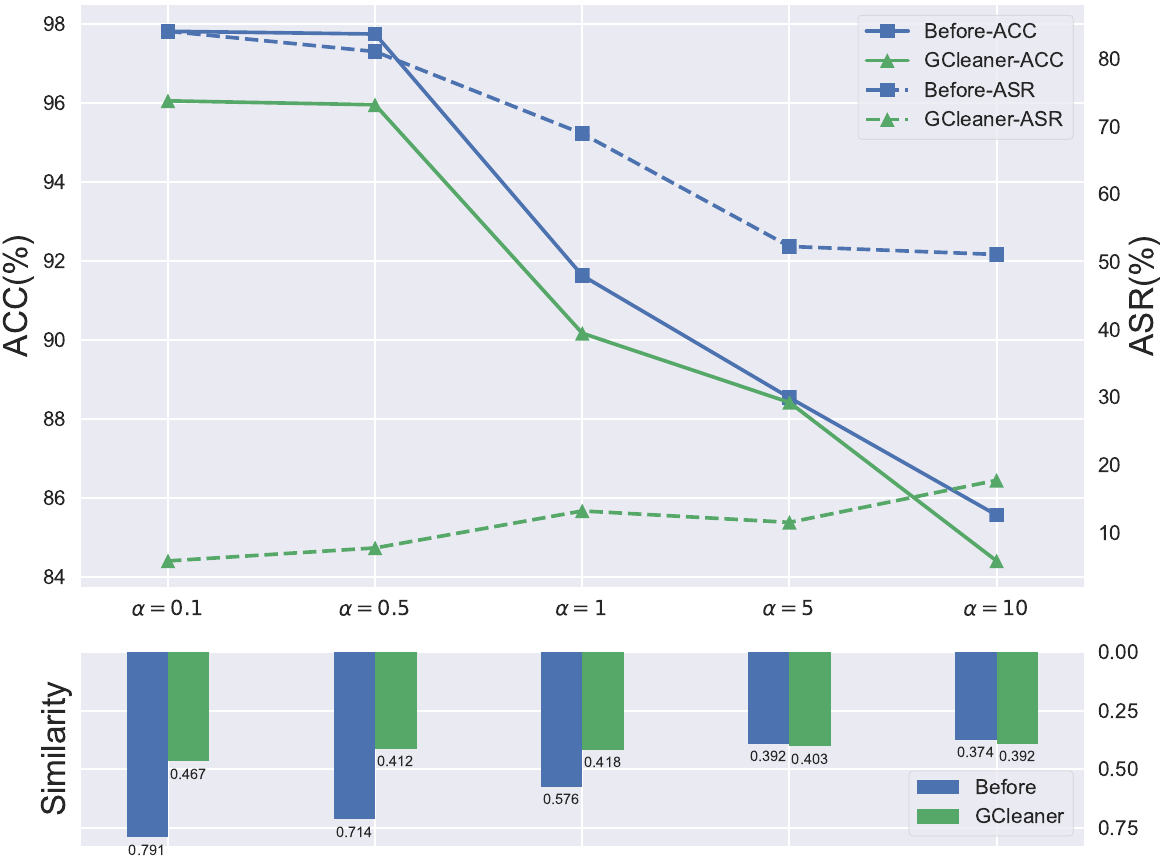}
	\caption{\sysname under adaptive attacks.}
\label{F9}
%\vspace{-3mm}
\end{figure}

\subsubsection{Impact of Explanation Methods}
\label{A.5}

We evaluated the impact of node importance scores (attention maps) generated by different explanation methods on our approach, specifically focusing on the variations of $\mathcal{L}_{explain}$ in Eq. \ref{EQ14}. Furthermore, we investigate the roles of the parameter $\lambda$ and the contribution of each loss component in Appendix \ref{A.7}. Our investigation focused on two primary categories: perturbation-based methods and gradient-based methods, namely GNNExplainer\cite{ying2019gnnexplainer}, IntegratedGradients (IGradients)\cite{sundararajan2017axiomatic}, and Guided BP \cite{baldassarre2019explainability}. The experimental results shown in Table \ref{T6} demonstrate that gradient-based methods outperform perturbation-based methods, with our approach achieving the best results. This is because perturbation-based methods infer feature importance by altering the input and observing changes in the output, which may not directly reflect the model's attention to backdoor nodes.

In contrast, gradient-based methods offer a more direct and reliable measure of the model’s internal response to input features. In the context of a backdoor attack, where an attacker embeds a specific trigger pattern into graphs, attention maps derived from gradient-based methods can precisely reveal the regions of the image that the model focuses on. This precise localization of the model’s attention is crucial for achieving finer-grained unlearning, as it allows for targeted mitigation of the backdoor effect without adversely affecting the model’s performance on benign inputs. Furthermore, our approach enhances this process by aggregating attention across multiple layers of the model, rather than relying on a single layer’s perspective. This multi-layer aggregation ensures a more holistic understanding of the model’s focus, resulting in a more effective defense.

\subsubsection{Adaptive Attacks}

Suppose the attacker trains the backdoored model to have different representations for backdoored samples to break the assumption of the proposed trigger recovery technique, rendering our defense ineffective. The attacker can achieve this by increasing $\alpha$ causing the backdoor-embedded samples to possess distinct feature representations. Therefore, the objective of the adaptive attack is as follows:
\begin{equation}
	\begin{split}
     \arg\min\mathcal{L}_{adapt}=\mathcal{L}_{atk}-\alpha\cdot\mathcal{L}_{p,q},\quad\alpha>0
	\end{split}
	\label{EQ16}
\end{equation}
where larger $\alpha$ means less similarity. We conducted experiments on the AIDS dataset. As shown in Figure \ref{F9}, with the increases of $\alpha$, it is impossible to achieve high ASR while maintaining ACC. This is because it aims to push poisoned samples away from each other and hence away from the samples from the target class, which contradicts the objective of the attack. In this scenario, the recovered backdoor trigger degrades into an adversarial perturbation that causes the model to output highly similar features. Although forgetting this perturbation cannot completely eliminate the backdoor features, it can enhance the robustness of the model to a certain extent.

\section{Related Work}
\subsection{Backdoor Attack and Defense on DNNs}
\label{sec: Backdoor Attack and Defense}
%Backdoor attack is a potent threat within Deep Neural Network (DNN) model training, manifesting as a simple yet formidable attack pattern. This technique jeopardizes the integrity of the training process by either contaminating training samples or exerting control over model parameters, ultimately encroaching upon subsequent tasks. Distinguished by its reliance on distinctive trigger patterns, backdoor attacks establish a robust correlation between the trigger and the desired target label, leading to the coercive alteration of the model's behavior and the formation of a surreptitious backdoor model. Importantly, this attack pattern exerts no influence on the model's accuracy with clean samples, while malicious objectives can only be realized when the backdoor model engages with tainted inputs, underscoring the calculated and precise nature of the adversary's intent.

In DNNs, backdoor attack is a potent threat, which can jeopardize the training process's integrity by contaminating training samples or exerting control over model parameters. Gu et al. \cite{Gu2019BadNetsEvaluating} initiated the exploratory endeavor into launching backdoor attacks (Badnets), employing pixel patches as triggers for activating covert manipulations. Subsequent studies have directed efforts towards enhancing the stealthy of these attacks, which can be divided into two categories. \ding{182}\textit{Invisible attacks}: numerous studies \cite{chen2017targeted,Li2021InvisibleBackdoor,Zhong2020BackdoorEmbedding} have innovated by introducing imperceptible alterations as the trigger mechanism for backdoor attacks. This subtlety is mainly achieved by minimizing the pixel-level discrepancies between the original and manipulated images \cite{zhang2024badcleaner}. To augment the stealth aspect of the attack, several approaches \cite{Doan2021BackdoorAttack,Ren2021SimtrojanStealthy,Zhao2022DEFEATDeep} have focused on maintaining the consistency in the latent space representation between the pristine and altered images by finessing the training loss function to integrate the backdoor into the model seamlessly. \ding{183}\textit{Natural attacks}: alternative approaches suggest altering the stylistic elements of the imagery to serve as the trigger, aimed at ensuring the images retain their natural appearance, thereby reducing suspicion. The creation of such natural triggers leverages diverse phenomena and technological applications such as natural attributes \cite{Liu2020ReflectionBackdoor} and generative adversarial network (GAN) \cite{Cheng2021DeepFeature}. 
Many studies \cite{chen2017targeted,Li2021InvisibleBackdoor,Liu2019ABSScanning,Liu2020ReflectionBackdoor,Nguyen2021WaNetImperceptible,Zhong2020BackdoorEmbedding} focus on stealthiness but neglect backdoor resilience. To address this, researchers have explored robust backdoor mechanisms, such as data augmentation for poisoned samples \cite{li2020rethinking,Zhang2022PoisonInk} and feature consistency training \cite{Xue2023CompressionresistantBackdoor,Wan2020FeatureConsistency} to reduce the impact of compression on triggered samples.

Extensive efforts in research have been devoted to addressing the impact of backdoors and cleansing compromised models. For example, NeuralCleanse (NC) employed reverse engineering to assess trigger effects and identified backdoors based on a specific threshold \cite{Wang2019NeuralCleanse}. Specifically, techniques for reverse trigger identification precisely locate neuron activation areas and establish filters to detect and eliminate backdoor neurons. SAGE introduced a top-down attention distillation mechanism that utilizes benign shallow layers to direct the mitigation of harmful deep layers, enhancing defensive capabilities through normalization and adaptive learning rate adjustments \cite{Gong2023RedeemMyself}. RAB offered a robust smoothing training algorithm to certify backdoor defense robustness without relying on noise distribution sampling \cite{Weber2023RABProvable}. In federated learning, BayBFed surpassed previous defense mechanisms by leveraging hierarchical Beta-Bernoulli and CRP-Jensen processes to calculate the probability of client updates for identifying and filtering malicious updates \cite{Kumari2023BayBFedBayesian}. However, the utilization of BayBFed may lead to a reduction in model performance. FLPurifier addresses this concern by segregating the model into a feature extractor and a classifier, disrupting the trigger pattern between backdoors and target labels before federated aggregation \cite{zhang2024flpurifier}.

%Research on backdoor impacts and model cleansing has led to several notable approaches. NeuralCleanse (NC) uses reverse engineering to identify backdoors by analyzing trigger effects and neuron activation areas \cite{Wang2019NeuralCleanse}. SAGE enhances defense through top-down attention distillation, guiding harmful deep layers with benign shallow ones and employing normalization and adaptive learning rates \cite{Gong2023RedeemMyself}. RAB provides robust backdoor defense with a smoothing training algorithm, avoiding noise distribution sampling \cite{Weber2023RABProvable}. In federated learning, BayBFed improves defenses by using hierarchical processes to filter malicious updates, though it may reduce model performance \cite{Kumari2023BayBFedBayesian}. FLPurifier mitigates this issue by separating the model into a feature extractor and classifier, disrupting backdoor-trigger patterns before federated aggregation \cite{zhang2024flpurifier}.

%\vspace{-2mm}

\subsection{Backdoor Attack and Defense on GNNs}
\label{sec: Backdoor Attack and Defense on GNNs}
Numerous studies have shown that backdoor attacks are effective in GNNs using different attack methods. Based on the different ways of generating triggers in graph backdoor attacks, existing work can be categorized as follows: 

\ding{182} \textbf{Tampering with node features}. %Xu et al. \cite{EXPBA} applied two powerful GNN interpretability methods to select the optimal node positions for trigger implantation and explore the effectiveness of backdoor attacks on node classification tasks. The method of seeking optimal nodes requires traversing the nodes, which is a time-consuming process for attackers. To address this limitation, Chen et al. \cite{chen2022general} used edge explanation and feature explanation methods to ensure that non-targeted nodes are not affected, achieving effective attacks on arbitrary nodes. Similarly, Xu et al. \cite{EXP2} quantitatively analyzed the differences between the most important or least important nodes, further deepening the understanding of backdoor attack behavior in GNNs. Dai et al. \cite{transferable} presented transferred semantic backdoors, assigning triggers to a specific class of nodes in the dataset, thereby evading detection by defenders. 
Xu et al. \cite{EXPBA} utilized GNN interpretability methods to identify optimal node positions for implanting triggers and assessed backdoor attacks on node classification tasks, though this node-traversal approach is time-consuming. To mitigate this, Chen et al. \cite{chen2022general} employed edge and feature explanation methods to target arbitrary nodes without affecting non-targeted ones. %Xu et al. \cite{EXP2} further analyzed the impact of node importance on backdoor attacks, while Dai et al. \cite{transferable} introduced transferred semantic backdoors, assigning triggers to specific node classes to evade detection.
Similarly, Xu et al. \cite{EXP2} quantitatively analyzed the differences between the most important or least important nodes, further deepening the understanding of backdoor attack behavior in GNNs. Dai et al. \cite{transferable} presented transferred semantic backdoors, assigning triggers to a specific class of nodes in the dataset, thereby evading detection by defenders. 

\ding{183} \textbf{Disrupting graph topology}. Xi et al. \cite{GAT} designed GTA, where triggers are defined as dynamically adjustable subgraphs. They optimize the triggers based on the topology of different graphs, achieving scalable attacks while ensuring independence from downstream tasks. Similarly, Zheng et al. \cite{motifBA} rethought triggers from the perspective of motifs, explaining the relationship between trigger structure and attack effectiveness. They found that the frequency of appearance of trigger subgraphs can influence the attack performance of the backdoor model. While subgraph embedding methods achieve effective attacks, they are too conspicuous for model detectors, resulting in weak attack evasion. Chen et al. \cite{neighboring} proposed neighboring backdoors, where triggers are designed as individual nodes connected to target nodes, forming the trigger pattern. If nodes are not connected, the model functions normally. The aforementioned attack methods introduce outputs with obviously incorrect labels, making them easily filtered in anomaly detection. To address this, Xu et al. \cite{clean} explored clean-label backdoor attacks, where the poisoned inputs generated have consistent labels with clean inputs, resulting in high attack success rates during testing. 
%Xi et al. \cite{GAT} developed GTA, using dynamically adjustable subgraphs as triggers to create scalable attacks independent of downstream tasks. Zheng et al. \cite{motifBA} explored triggers from a motif perspective, showing that the frequency of trigger subgraphs impacts attack effectiveness. Chen et al. \cite{neighboring} proposed neighboring backdoors with triggers as individual nodes linked to target nodes, ensuring normal model function if nodes are disconnected. However, these methods often produce outputs with incorrect labels, making them detectable. To overcome this, Xu et al. \cite{clean} introduced clean-label backdoor attacks, where poisoned inputs share consistent labels with clean ones, improving attack success rates during testing.

\ding{184} \textbf{Node information alteration}. Graph Contrastive Backdoor Attacks (GCBA) \cite{contras} introduced the first backdoor attack for graph contrastive learning (GCL). They devised different backdoor strategies for different stages of the GCL pipeline: data poisoning, post-pretraining encoder backdoor injection, and natural backdoor. Besides, Unnoticeable Graph Backdoor Attack (UGBA) \cite{dai2023unnoticeable} considers how to achieve more covert attacks with limited attack budgets. It intentionally selects nodes carrying the backdoor during the poisoning stage to save the budget and deploys adaptive triggers for stealthy attacks. However, defending against backdoor attacks in GNNs is largely unexplored, which motivates us to design the first effective graph backdoor defense mechanism in our paper.
%Graph Contrastive Backdoor Attacks (GCBA) \cite{contras} introduced the first backdoor attack for graph contrastive learning (GCL), employing various strategies across different stages of the GCL pipeline, including data poisoning and post-pretraining encoder injection. Unnoticeable Graph Backdoor Attack (UGBA) \cite{dai2023unnoticeable} aimed to achieve covert attacks with limited budgets by selectively targeting nodes for backdoor insertion and using adaptive triggers. The limited exploration of defenses against backdoor attacks in GNNs has prompted the development of the first effective graph backdoor defense mechanism in our study.

\section{Conclusion \& Discussion}

In this paper, we propose \sysname, the first backdoor mitigation method on GNNs. \sysname can effectively mitigate the negative influences by simply reversing the backdoor learning procedure. Specifically, \sysname consists of two primary modules, named trigger recovery module and backdoor unlearning module. In the first step, \sysname employs explanation methods to identify optimal trigger locations, facilitating the search of universal and hard backdoor triggers in the feature space of the backdoored model through maximal similarity. With the recovered triggers, \sysname leveraged them to unlearn the backdoor patterns embedded in a GNN model. That is, \sysname employs the backdoored model as the student model, and the fine-tuned and frozen model as the teacher model. Through the teacher's guidance of the student's paradigm combined with intermediate explainable knowledge, \sysname can successfully achieve the backdoor unlearning. Extensive experiments validate that \sysname is effective against known backdoor attacks in GNNs with negligible performance degradation and outperforms the SOTA defense methods.

\textbf{Limitations \& Future work.} One limitation of our study pertains to the requisite utilization of a fraction of the pristine dataset sourced from the original dataset to recover trigger subgraphs and execute the unlearning procedure. While our investigation has substantiated that, even under extreme circumstances, leveraging auxiliary public datasets with distributions differing from the original data can still manifest effective backdoor defense, an associated degradation in performance is observed. Furthermore, our focus lies predominantly on backdoor defenses on subgraph classification. The transference of existing defense strategies to node prediction scenarios remains a pressing issue demanding resolution. In forthcoming endeavors, we aim to delve deeper into scenarios where no dataset interaction is feasible, achieving backdoor alleviation for graphs while also exploring a universal backdoor defense framework tailored to diverse graph tasks.

\iffalse
%-------------------------------------------------------------------------------
\section*{Acknowledgments}
%-------------------------------------------------------------------------------

The USENIX latex style is old and very tired, which is why
there's no \textbackslash{}acks command for you to use when
acknowledging. Sorry.
\fi

\newpage
%-------------------------------------------------------------------------------
% \footnotesize
\bibliographystyle{IEEEtran}
\bibliography{main}

\newpage

\appendices
% \normalsize

\section{Details on Datasets}
\label{A.1}
\begin{itemize}
    \item \textbf{Bitcoin}: It is a dataset for detecting fraudulent Bitcoin transaction. Each graph in the dataset represents a transaction, with a transaction and the transactions that have Bitcoin flow with it as nodes, and the flow of Bitcoin currency is represented by edges. Each graph is labeled as 0 or 1, corresponding to illicit and legitimate transactions, respectively.
    \item \textbf{Fingerprint}: Fingerprint is a collection of fingerprints formatted as graph structures from the NIST-4 database, which consists of 4,000 grayscale images of fingerprints with class labels according to the five most common classes of the Galton-Henry classification scheme.
    \item \textbf{AIDS}: It is a collection of anonymized medical records from patients diagnosed with acquired immune deficiency syndrome (AIDS). This dataset contains information on various demographic, clinical, and laboratory variables, such as age, sex, CD4 cell count, viral load, and the presence of opportunistic infections.
    \item \textbf{COLLAB}: It is a scientific collaboration dataset. Each graph in the dataset represents a researcher's ego network, where the researcher and their collaborators are nodes, and collaboration between researchers is indicated by edges. The ego network of a researcher is classified into three labels: High Energy Physics, Condensed Matter Physics, and Astro Physics, representing their respective fields of study.
\end{itemize}

\section{Details on Attack Methods}
\label{A.2}
\begin{itemize}
    \item \textbf{Subgraph-based Backdoor (Sub-BA)} \cite{badsub}: Sub-BA involves injecting a subgraph trigger into a graph. In our experiments, we generate the subgraph trigger using the Erdős-Rényi (ER) model \cite{ER}. The poisoned nodes are randomly selected from the node set of the graph.
    \item \textbf{GTA} \cite{GAT}: To execute the GTA attack, we train a topology generator and a feature generator separately for five epochs. These generators enable us to generate a range of candidate backdoor triggers. We subsequently solve a bi-level optimization problem iteratively to poison the victim graphs with the backdoor triggers. For all experiments, we set the number of epochs as 20 for the bi-level optimization process.
	\item \textbf{Explainability-based Attack (Exp-BA)} \cite{EXPBA}: Exp-BA initially employs the Graph-Explainer \cite{ying2019gnnexplainer} to generate a node importance matrix for each victim graph using a pre-trained model. Subsequently, it replaces the nodes with the top-k highest importance values with the subgraph trigger, where $k$ represents the number of poisoned nodes.
    \item \textbf{Motif-Backdoor (Motif-BA)} \cite{motifBA}: Motif-BA first obtains the distribution of the motif through the motif extraction tool and analyzes it to select the suitable motif as the trigger. Then, the trigger injection position is determined by using network importance metrics, shadow models, and target node-dropping strategies. Finally, it injects the trigger into benign graphs and learns a backdoored model on them.

\end{itemize}

\section{Comparison with Detection Methods}
\label{A.3}

Table \ref{T7} presents a comparative analysis between \sysname and two detection methods across four backdoor attacks. ED and XGBG, as detectors, have the capability to access the entire training set, detect, and subsequently exclude backdoor samples. Despite achieving noteworthy detection accuracy, a residual presence of a small set of backdoor samples within the data remains possible, especially when confronted with optimizable triggers like GTA. Even a scarce number of backdoor samples are adequate to orchestrate a successful backdoor attack. Furthermore, detection techniques are unable to erase the backdoor characteristics embedded within pre-trained backdoored models. Consequently, \sysname exhibits superior adaptability in addressing these challenges.

\begin{table}[h]
 \caption{\small Performance of \sysname under different number of clients.}
 \vspace{1mm}
  \centering
  \footnotesize
  \renewcommand\tabcolsep{7pt}
  \renewcommand\arraystretch{1.2}
\begin{tabular}{c|cccccc}
\hline
\multirow{2}{*}{\textbf{Attack}} & \multicolumn{2}{c}{\textbf{ED}} & \multicolumn{2}{c}{\textbf{XGBD}} & \multicolumn{2}{c}{\textbf{Ours}} \\ \cline{2-7} 
                                 & \textbf{ASR}   & \textbf{ACC}   & \textbf{ASR}    & \textbf{ACC}    & \textbf{ASR}    & \textbf{ACC}    \\ \hline
\textbf{Sub-BA}                  & 26.64          & 90.37          & 17.18           & 89.51           & \cellcolor{lightblue}6.64            & \cellcolor{lightblue}95.37           \\
\textbf{GTA}                     & 52.04          & 89.64          & 32.28           & 92.33           & \cellcolor{lightblue}2.04            & \cellcolor{lightblue}94.64           \\
\textbf{Exp-BA}                  & 16.81          & 92.12          & 9.26            & 93.35           & \cellcolor{lightblue}6.81            & \cellcolor{lightblue}96.12           \\
\textbf{Motif-BA}                & 24.08          & 87.36          & 15.32           & 92.94           & \cellcolor{lightblue}4.08            & \cellcolor{lightblue}95.36           \\ \hline
\end{tabular}
  \label{T7}
\end{table}

\begin{figure*}[t]
	\centering
 \subfigure[Backdoored model with Sub-BA.]{
		\begin{minipage}[t]{0.235\linewidth}
			\centering
			\includegraphics[width=1\linewidth]{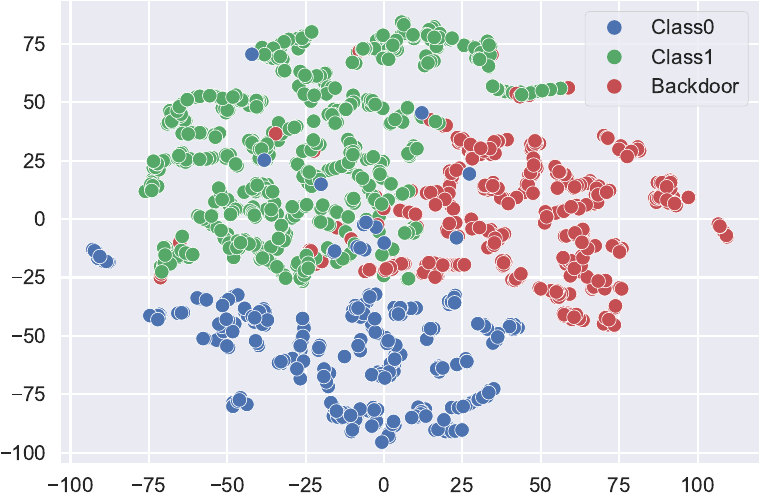}
			\label{F11a}
	\end{minipage}}
	\subfigure[Backdoored model with GTA.]{
		\begin{minipage}[t]{0.235\linewidth}
			\centering
			\includegraphics[width=1\linewidth]{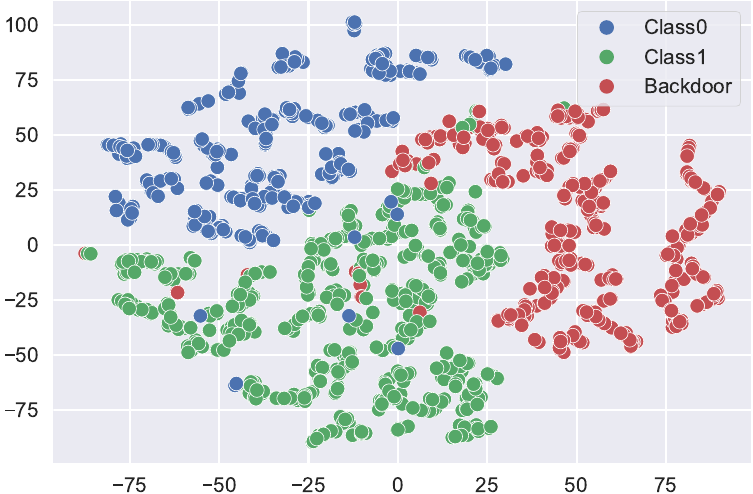}
			\label{F11b}
	\end{minipage}}
	 \subfigure[Backdoored model with Exp-BA.]{
		\begin{minipage}[t]{0.235\linewidth}
			\centering
			\includegraphics[width=1\linewidth]{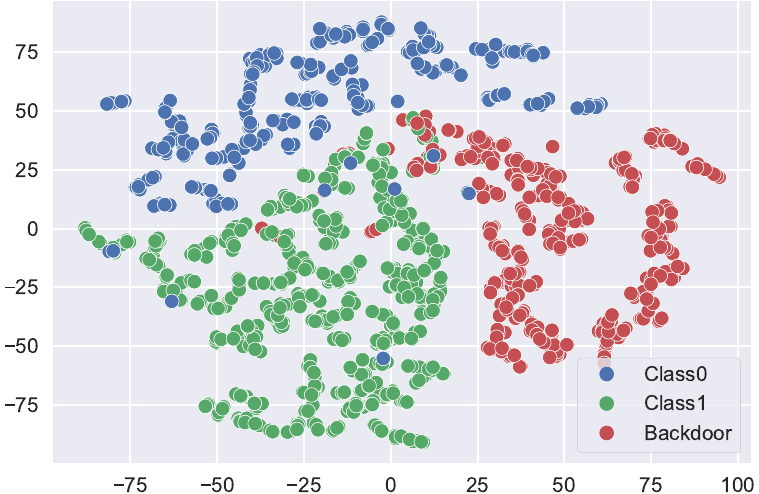}
			\label{F11c}
	\end{minipage}}
  \subfigure[Backdoored model with Motif-BA.]{
		\begin{minipage}[t]{0.235\linewidth}
			\centering
			\includegraphics[width=1\linewidth]{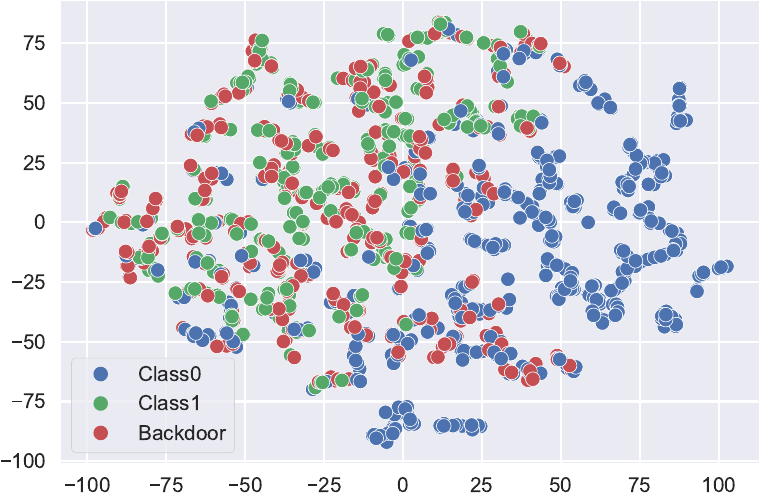}
			\label{F11d}
	\end{minipage}}
 \subfigure[Unlearning model with Sub-BA.]{
		\begin{minipage}[t]{0.235\linewidth}
			\includegraphics[width=1\linewidth]{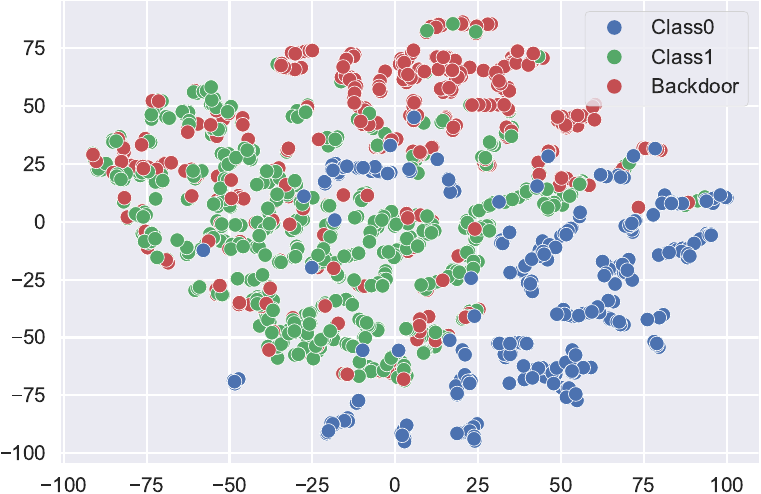}
			\label{F11e}
	\end{minipage}}
	\subfigure[Unlearning model with GTA.]{
		\begin{minipage}[t]{0.235\linewidth}
			\includegraphics[width=1\linewidth]{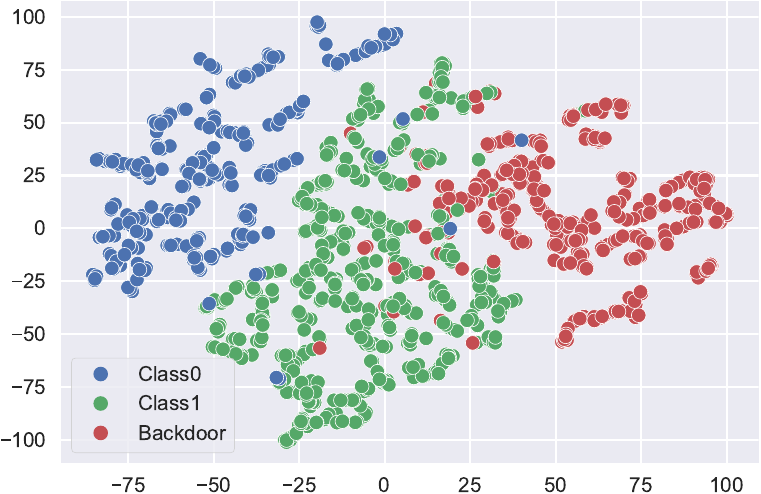}
			\label{F11f}
	\end{minipage}}
  \subfigure[Unlearning model with Exp-BA.]{
		\begin{minipage}[t]{0.235\linewidth}
			\includegraphics[width=1\linewidth]{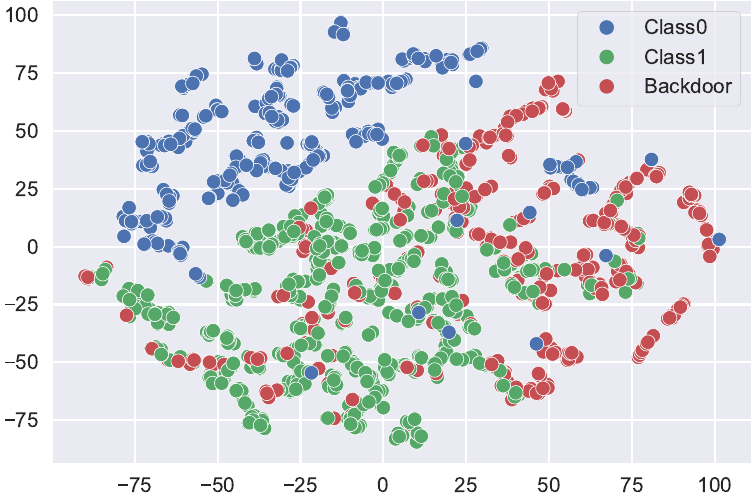}
			\label{F11g}
	\end{minipage}}
	\subfigure[Unlearning model with Motif-BA.]{
		\begin{minipage}[t]{0.235\linewidth}
			\includegraphics[width=1\linewidth]{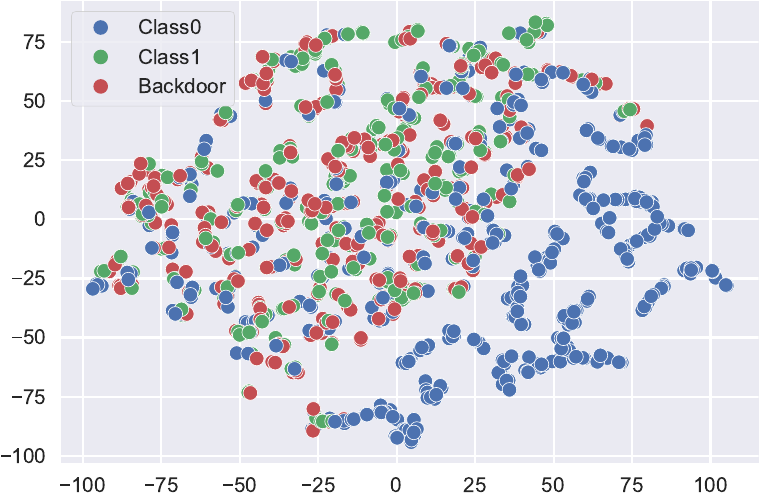}
			\label{F11h}
	\end{minipage}}
	\caption{Visualizations of feature embedding of the backdoored model and unlearning model.}
	\label{F11}
	%\vspace{-2mm}
\end{figure*}

\section{Explanation Visualizations}
\label{A.6}
Figure \ref{F11} displays the visualization of the distribution in the feature space of the model with/without \sysname against four attacks. The experimental findings indicate that following the process of unlearning, the features of the embedded backdoor trigger samples realign themselves within their original class, distinctly distant from the target class. Specifically, we visualize the distribution in the feature space of the model under the above attacks with/without \sysname to further understand the advantages of \sysname. More specifically, we sample 1800 samples from AIDS, of which 600 are from class 0 and 1200 are from class 1 (half of which are embedded triggers). Then, the data are fed to Sub-BA attacked backdoored models and output the feature embeddings. These features are processed by the T-SNE algorithm and visualized. From Figure \ref{F11}, we can see that the feature representations of benign samples from the same category form an individual cluster, while the poisoned samples form a new cluster (in red color). However, the clusters of poisoned samples are damaged, all the poisoned samples afresh assemble with the benign samples from the same category. This means that \sysname can effectively break the cluster of backdoor features to unlearn the backdoor feature. Moreover, the benign samples are located in their own cluster in a relatively concentrated manner, where the distance between samples within clusters is very close meanwhile there is a significant boundary between clusters, illustrating that the model shows excellent ability on the main classify task.  %Please refer to the Appendix \ref{A.6} for other attacks.
%\vspace{-3mm}

\iffalse
\begin{figure}
	\centering
	\subfigure[Backdoored model with Sub-BA.]{
		\begin{minipage}[t]{0.48\linewidth}
			\centering
			\includegraphics[width=1\linewidth]{F8a.pdf}
			\label{F8a}
	\end{minipage}}
	\subfigure[Unlearning model with Sub-BA.]{
		\begin{minipage}[t]{0.48\linewidth}
			\includegraphics[width=1\linewidth]{F8b.pdf}
			\label{F8b}
	\end{minipage}}%
	\caption{Visualizations of feature embedding of the backdoored model and unlearning model.}
	\label{F8}
	%\vspace{-3mm}
\end{figure}
\fi

\section{Ablation Studies}
\label{A.7}

\begin{table}[]
 \caption{\small Performance of \sysname under different number of recovered trigger nodes.}
  \vspace{1mm}
  \centering
  \footnotesize
  \renewcommand\tabcolsep{5pt}
  \renewcommand\arraystretch{1}
\begin{tabular}{c|c|c|c|c|c|c|c}
\hline
\textit{\textbf{Num}} & \textbf{n=1} & \textbf{n=2} & \textbf{n=3} & \textbf{n=4} & \textbf{n=5} & \textbf{n=6} & \textbf{n=7} \\ \hline
\textit{\textbf{ASR}} & 23.13        & 13.51        & 6.64         & 4.28         & 2.14         & 1.17         & 0.58         \\ \hline
\textit{\textbf{ACC}} & 97.82        & 96.56        & 95.37        & 95.50        & 90.95        & 86.37        & 80.64        \\ \hline
\end{tabular}
\label{T8}
\end{table}
 % \vspace{-3mm}

\textbf{Number of recovery triggers.}
During the trigger subgraph recovery phase, the defender remains unaware of the specific number of nodes in the trigger embedded in the backdoor. To address this uncertainty, we employ a predetermined number to recreate the trigger, ensuring it encapsulates features that closely resemble or are equivalent to the original trigger subgraph. We systematically investigate the influence of varying the preset number of nodes on the ultimate efficacy of defense mechanisms. Experiments were conducted on the AIDS dataset against Sub-BA, with results summarized in Table \ref{T8}.

The analysis reveals that with smaller trigger subgraphs, the complete abstraction of the backdoor trigger features learned by the model becomes challenging, thereby constraining the effectiveness of backdoor defense mechanisms. Conversely, employing excessively large presets allows for a comprehensive reconstruction of trigger features but risks impairing the primary task during the forgetting of backdoor features. Consequently, the judicious selection of an appropriate number of preset trigger nodes emerges as a critical factor in optimizing trigger restoration strategies.

\begin{figure}
\centering
\includegraphics[width=0.85\linewidth]{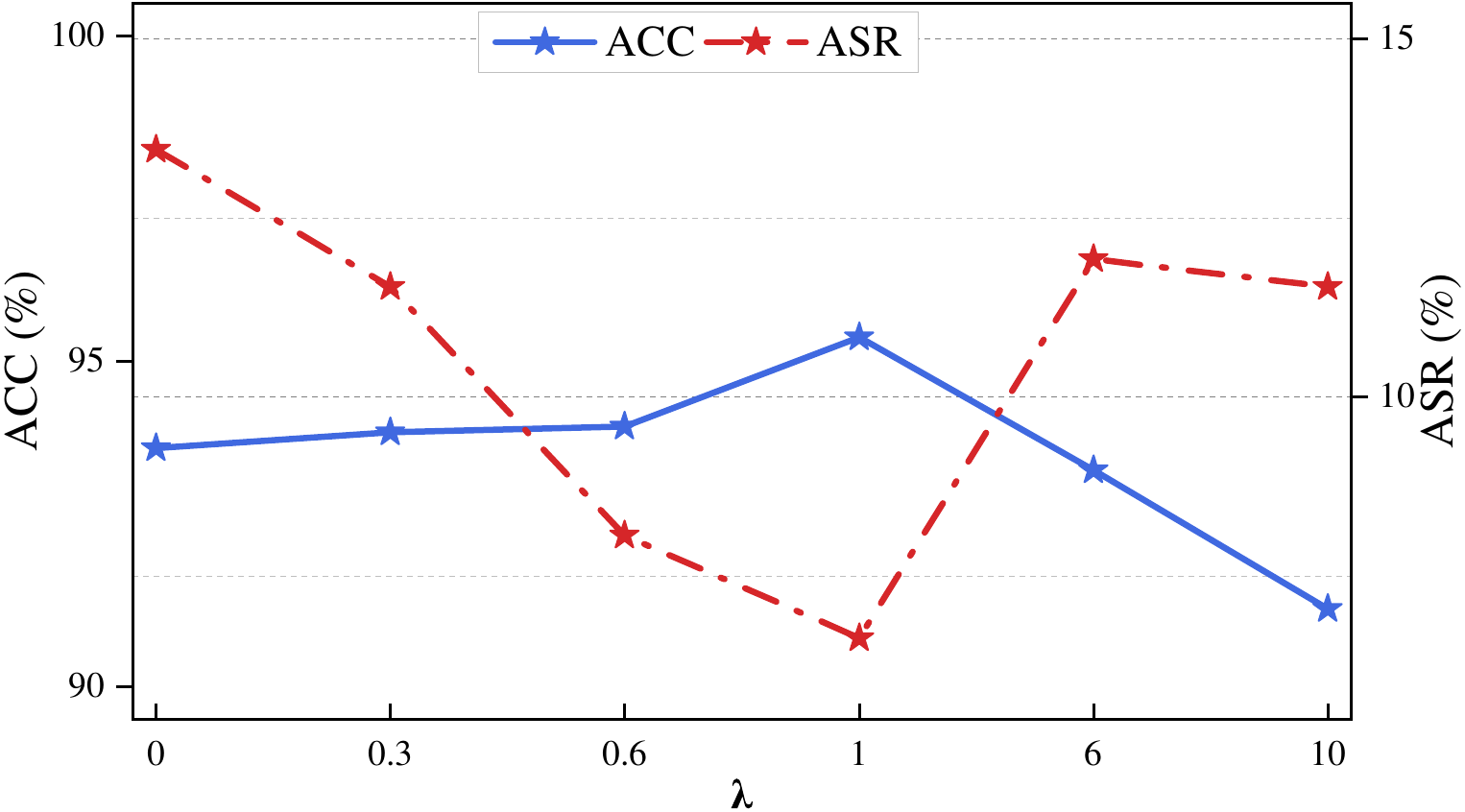}
\caption{Impact of loss term $\lambda$.}
\label{F12}
%\vspace{-5mm}
\end{figure}

\begin{figure*}[t]
	\centering
	\subfigure[ASR on Bitcoin dataset.]{
		\begin{minipage}[t]{0.235\linewidth}
			\centering
			\includegraphics[width=1\linewidth]{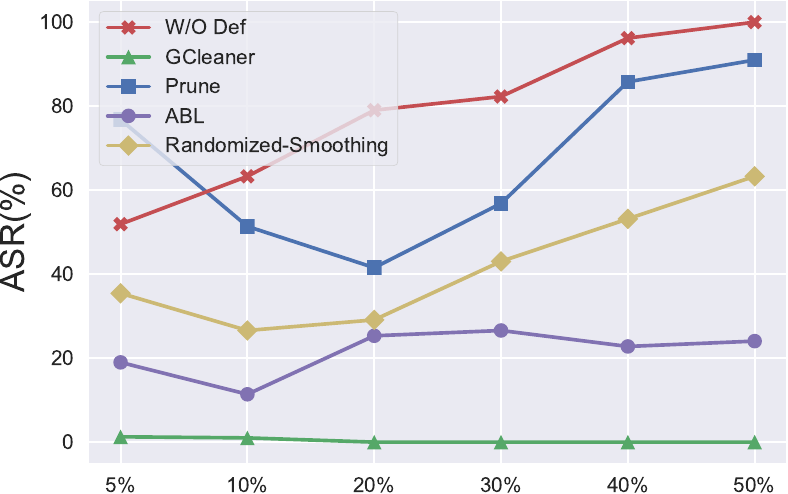}
			\label{F10a}
	\end{minipage}}
	\subfigure[ASR on Fingerprint dataset.]{
		\begin{minipage}[t]{0.235\linewidth}
			\includegraphics[width=1\linewidth]{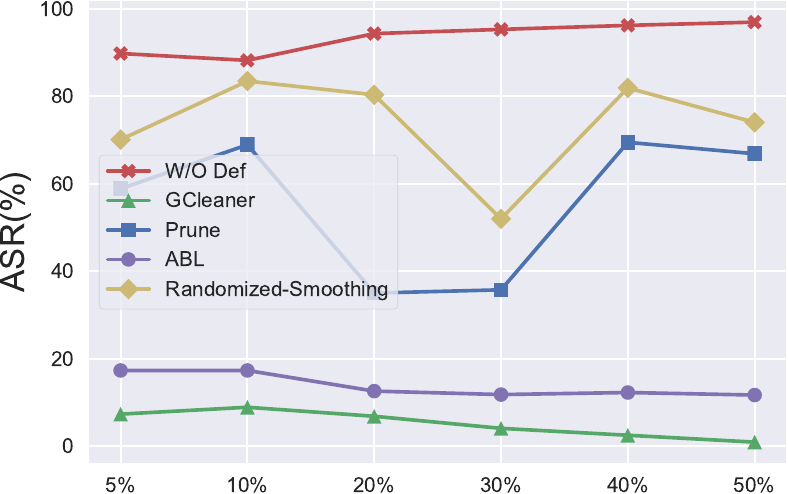}
			\label{F10b}
	\end{minipage}}
	\subfigure[ASR on AIDS dataset.]{
		\begin{minipage}[t]{0.235\linewidth}
			\centering
			\includegraphics[width=1\linewidth]{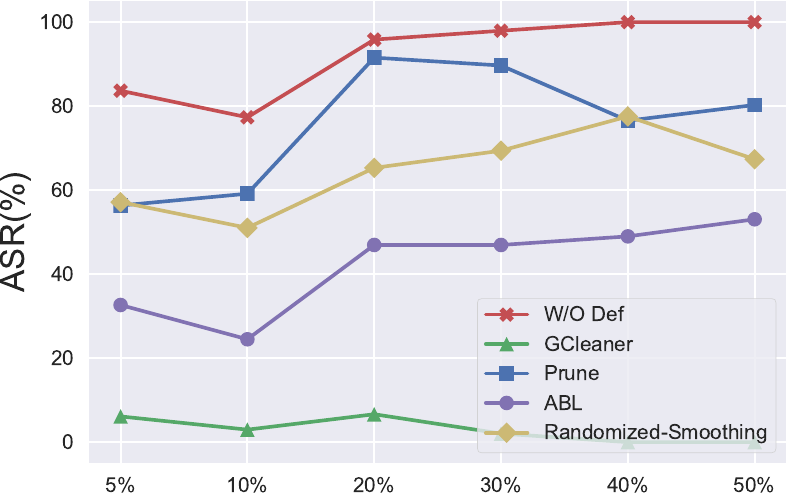}
			\label{F10c}
	\end{minipage}}
	\subfigure[ASR on COLLAB dataset.]{
		\begin{minipage}[t]{0.235\linewidth}
			\includegraphics[width=1\linewidth]{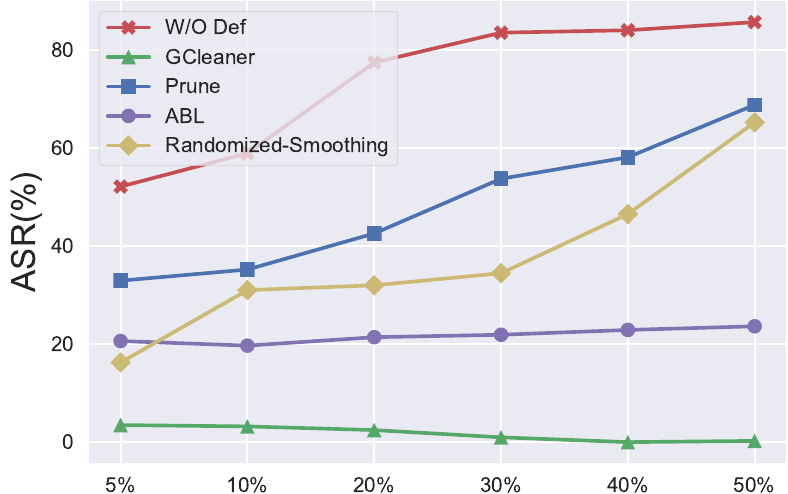}
			\label{F10d}
	\end{minipage}}%
 \\
 \subfigure[ACC on Bitcoin dataset.]{
		\begin{minipage}[t]{0.235\linewidth}
			\centering
			\includegraphics[width=1\linewidth]{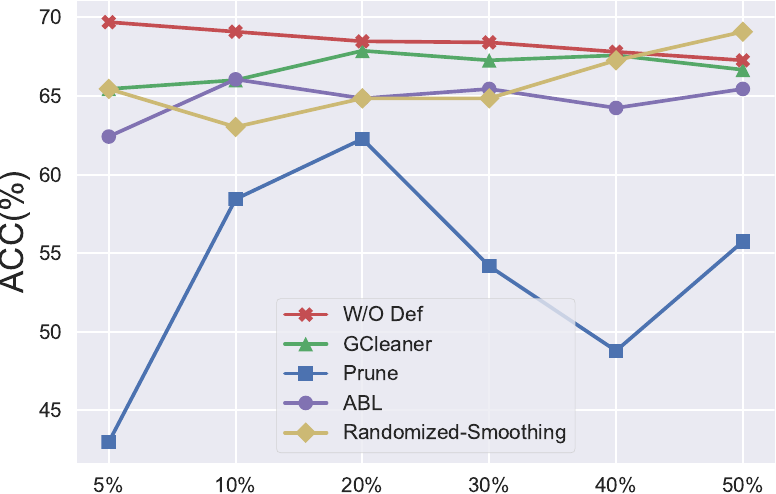}
			\label{F10e}
	\end{minipage}}
	\subfigure[ACC on Fingerprint dataset.]{
		\begin{minipage}[t]{0.235\linewidth}
			\includegraphics[width=1\linewidth]{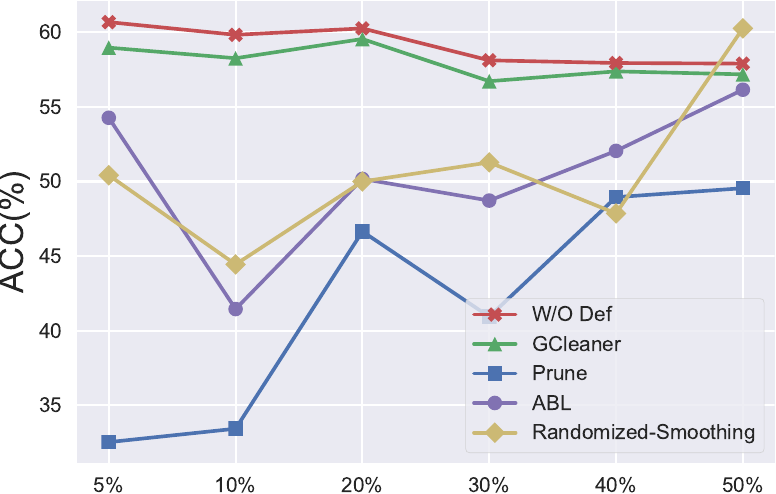}
			\label{F10f}
	\end{minipage}}
	\subfigure[ACC on AIDS dataset.]{
		\begin{minipage}[t]{0.235\linewidth}
			\centering
			\includegraphics[width=1\linewidth]{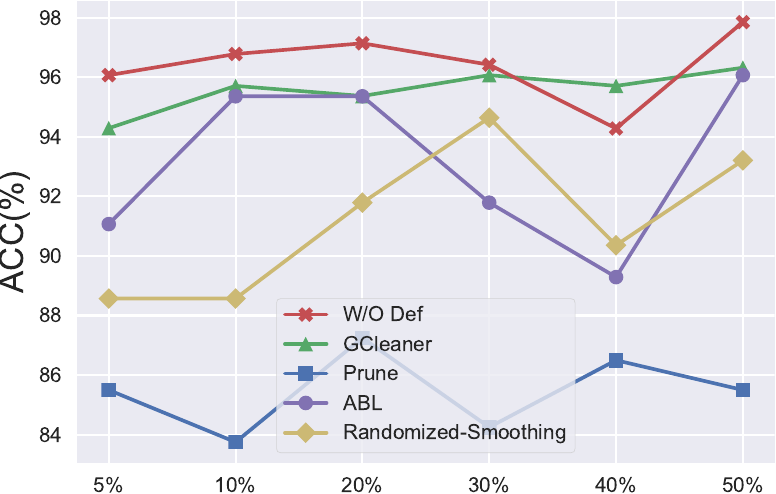}
			\label{F10g}
	\end{minipage}}
	\subfigure[ACC on COLLAB dataset.]{
		\begin{minipage}[t]{0.235\linewidth}
			\includegraphics[width=1\linewidth]{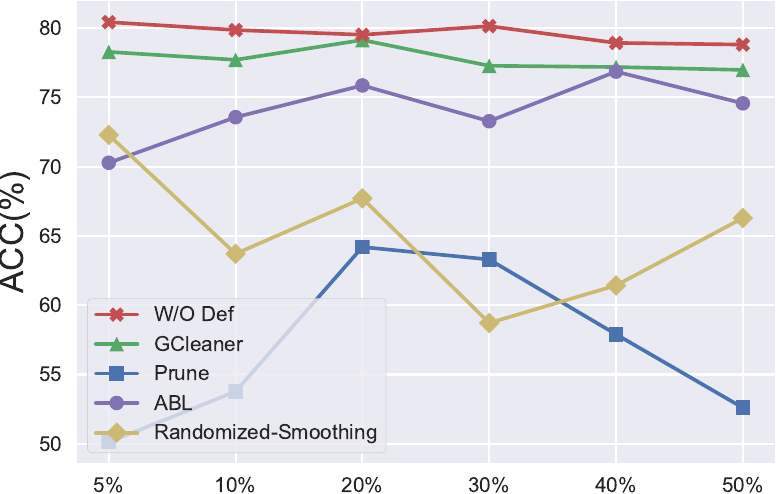}
			\label{F10h}
	\end{minipage}}%
	\caption{Impact of trigger size on four datasets.}
	\label{F10}
	%\vspace{-2mm}
\end{figure*}

\indent\textbf{Loss terms.}
The hyperparameter $\lambda$ holds significant importance in striking a balance between $\mathcal{L}_{unlearn}$ (the front part) and $\mathcal{L}_{explain}$ (the back part). Traditionally, $\lambda$ is a constant value derived through empirical parameter tuning. Recent investigations have introduced a linear decay mechanism to gradually anneal $\lambda$. This method facilitates a smooth transition from guided learning, reminiscent of learning under a teacher's supervision, towards self-regulated learning. This approach primarily addresses challenges arising from substantial disparities in size between teacher and student models.

In our study, both the teacher model and the student model maintain an identical size. Hence, our focus is solely on examining the ramifications of various fixed values of $\lambda$ within our methodology, noting that the default setting is $\lambda=1$. The outcomes are depicted in Figure \ref{F12}. Overall, the mitigation of backdoor effects demonstrates an enhancement as $\lambda$ increases, but excessive values can have the opposite effect. Optimal distillation results are achievable with a judiciously chosen moderate value of $\lambda$.

\begin{table}[t]
 \caption{\small Performance of \sysname with different components.}
  \vspace{1mm}
  \centering
  \footnotesize
  \renewcommand\tabcolsep{8pt}
  \renewcommand\arraystretch{1.2}
\begin{tabular}{ccc|cc}
\hline
$\mathcal{L}_{unlearn}^1$ & $\mathcal{L}_{unlearn}^2$ & $\mathcal{L}_{explain}$ & \textbf{ASR} & \textbf{ACC} \\ \hline
     \textcolor{red}{\ding{55}}                &          \textcolor{red}{\ding{55}}           &                    \textcolor{red}{\ding{55}}      & 95.84             & 97.14             \\ \hline
\textcolor{green}{\ding{51}}                     &           \textcolor{red}{\ding{55}}           &           \textcolor{red}{\ding{55}}               & 9.74            & 79.12             \\ \hline
\textcolor{green}{\ding{51}}                    & \textcolor{green}{\ding{51}}                    &                \textcolor{red}{\ding{55}}         & 13.01             & 92.41             \\ \hline
\textcolor{green}{\ding{51}}                    & \textcolor{green}{\ding{51}}                     & \textcolor{green}{\ding{51}}                         & 6.64             & 95.37             \\ \hline
\end{tabular}
  \label{T9}
 % \vspace{-3mm}
\end{table}

\textbf{Component Contributions.}
 Our unlearning losses are divided into two components: unlearning loss $\mathcal{L}_{unlearn}$ and intermediate explainable loss $\mathcal{L}_{explain}$, which the unlearning loss can be further separated into $\mathcal{L}_{unlearn}^1$ and $\mathcal{L}_{unlearn}^2$. We delve into the roles of each component to advance our understanding of the working mechanisms behind the unlearning process. Table \ref{T9} presents the contributions of different components. It is evident that when utilizing only $\mathcal{L}_{unlearn}^1$, while capable of achieving backdoor feature forgetting, there is a lack of maintenance in ACC. Noteworthy is our earlier approach where we employed the outputs of the teacher model and the student model separately to facilitate backdoor unlearning. However, we discovered that exclusively leveraging the components of the student model itself yields superior results, which is attributed to the continual improvement of the student model during optimization, akin to self-purify.

Utilizing $\mathcal{L}_{unlearn}^1$ and $\mathcal{L}_{unlearn}^2$ in tandem allows for the preservation of ACC while unlearning backdoor features. Indeed, the outcomes at this stage are already promising. The incorporation of explainable loss further enhances the overall performance, indicating a refined efficacy in our methodology.

\section{Impact of Trigger Size}
\label{A.4}

Figure \ref{F10} illustrates the impact of trigger size on \sysname across different datasets. The X-axis represents the trigger size, ranging from 5$\%$ to 50$\%$ of the original graphic size. Experimental findings indicate that between 5$\%$ and 30$\%$, the Attack Success Rate (ASR) significantly increases as the trigger size grows, with a subsequent slowdown in growth thereafter. For ABL, there is no discernible pattern in the effectiveness of backdoor defense concerning trigger size variation. In contrast, both Prune and Randomized-Smoothing display a decline in defense performance as the trigger size increases. The former leverages node pruning, where an increase in trigger nodes leads to a higher false positive rate, thereby disrupting the defense efficacy. The latter involves random sampling during training, and a larger trigger size results in a higher likelihood of sampling trigger nodes, consequently causing an elevation in ASR. In contrast, a larger trigger size benefits our backdoor restoration process and enhances the precision of backdoor forgetting, making defense somewhat easier. This facilitates more accurate backdoor identification, ultimately aiding in defense mechanisms.

% In addition, we validated the impact of attention maps (node importance scores) generated by different explanation methods as intermediate layer knowledge on our method, as shown in Appendix \ref{A.5}.
% \vspace{-3mm}

%%%%%%%%%%%%%%%%%%%%%%%%%%%%%%%%%%%%%%%%%%%%%%%%%%%%%%%%%%%%%%%%%%%%%%%%%%%%%%%%
\end{document}